\documentclass[showpacs,preprintnumbers,amsmath,amssymb,latexsym,array,enumerate,letter,11pt]{revtex4-1}
\usepackage{graphicx}
\usepackage{dcolumn}
\usepackage{bm}
\usepackage{amssymb}
\usepackage{epsfig}
\usepackage{slashed}
\usepackage{inputenc}

\newcommand{\half}{\frac{1}{2}}
\newcommand{\be}{\begin{equation}}
\newcommand{\ee}{\end{equation}}
\newcommand{\beq}{\begin{eqnarray}}
\newcommand{\eeq}{\end{eqnarray}} 
\newcommand\eqn[1]{\label{eq:#1}} 
\newcommand\eq[1]{eq. (\ref{eq:#1})} 
\newcommand{\vev}[1]{\langle #1 \rangle}

\newcommand{\GeV}{{\rm ~GeV }}
\newcommand{\TeV}{{\rm ~TeV }}

\newcommand{\CM}{{\cal M}}

\newcommand{\CY}{{\cal Y}}

\newcommand{\CL}{{\cal L}}

\newcommand{\mybar}[1]%
        {\kern  .6pt\overline{\kern - .6pt#1\kern - .6pt}\kern  .6pt}
\begin{document}

\title{Little Flavor: Heavy Leptons, $Z'$ and Higgs Phenomenology }
\author{Sichun Sun}
\email{sichun@uw.edu}

\affiliation{Institute for Nuclear Theory, Box 351550, Seattle, WA 98195-1550, USA}
\affiliation{Institute for Advanced Study, Hong Kong University of Science and Technology, Clear Water Bay, Hong Kong}

\date{\today}

\begin{abstract}
The Little Flavor model \cite{2013PhRvD..87l5036S} is a close cousin of the Little Higgs theory which aims to generate flavor structure around TeV scale. While the original Little Flavor only included the quark sector, here we build the lepton part of the Little Flavor model and explore its phenomenology. The model produces the neutrino mixing matrix and Majorana masses
of the Standard Model neutrinos through coupling to heavy lepton partners and Little Higgses. We combine the usual right-handed seesaw mechanism with global symmetry protection to suppress the Standard Model neutrino masses, and identify the TeV partners of leptons as right-handed Majorana neutrinos. The lepton masses and mixing matrix are calculated perturbatively in the theory. 

The TeV new 
gauge bosons have suppressed decay width in dilepton channels. Even assuming the Standard Model couplings, the branching ratios to normal dilepton channels are largely reduced in the model, to evade the bound from current $Z'$ search. It also opens up the new search channels for exotic gauge bosons, especially $Z' \rightarrow \slashed E_{t} + \text{multi } \ell+jets $. The multiple lepton partners will create new chain decay signals in flavor related processes in colliders, which also give rise to flavor anomalies. The lepton flavor violation process can be highly suppressed in charged lepton sector and happens only through neutrinos.
\end{abstract}

\maketitle

 \section{Introduction}

 The structure and origin of flavor remain mysteries in the Standard Model (SM). There are many models about flavor, although none of them seems particularly convincing. Moreover the majority of the models rely on high scale/short distance to explain the absence of observed electric dipole moments, the small values for the neutrino masses, and flavor changing rare decays such as $\mu \rightarrow 3e$. This is also part of the reason that most flavor models lack experimental proofs.

The Little/composite Higgs theory \cite{Kaplan:1983fs,Kaplan:1983sm,ArkaniHamed:2002qy,ArkaniHamed:2002qx,Schmaltz:2005ky,ArkaniHamed:2001nc} is an alternative to supersymmetry to address the hierarchy problem at low energy. The classes of little Higgs/composite Higgs types of theories now have generated numerous realistic extensions\cite{Marzocca:2012zn,Ecker:1989yg,Giudice:2007fh,Barbieri:2015lqa,Rattazzi:2000hs,Low:2014oga,Low:2015ogb,Cheng:2003ju,Low:2004xc,Low:2015ogb,Contino:2006qr,Anastasiou:2009rv,Medina:2007hz, Agashe:2004rs,Panico:2011pw,DeCurtis:2011yx,Giudice:2007fh,Grossman:1999ra,Barbieri:2015lqa,Gripaios:2009pe,Mrazek:2011iu,Hall:2001zb,Kubo:2001zc,Burdman:2002se,Scrucca:2003ra,Contino:2003ve,Panico:2005dh,Panico:2006em,Contino:2006qr} and phenomenological studies in colliders\cite{Vignaroli:2012sf,DeSimone:2012fs,Delaunay:2013iia,Gillioz:2013pba,Han:2003wu,Carena:2006jx,Matsumoto:2008fq,Contino:2015gdp,Huber:2000ie,Matsedonskyi:2012ym}. It predicts fewer new particles comparing to supersymmetry. Each fermion can have an exotic fermionic partner, Higgses have pseudo Goldstone bosons as partners; The Little/composite Higgs theory uses softly broken nonlinearly realized symmetries to protect the Higgs mass. 

The ultra-violet completion of the strong dynamics in little/composite Higgs theory can be directly related to the recent "second string revolution" in 00s. The Ads/CFT correspondence and Randall-Sundrum type of theories indicate that higher dimensional supersymmetric theories are dual to conformal/non-conformal theories in four dimensions through holography,e.g.\cite{Rattazzi:2000hs,McGuirk:2009am,Gherghetta:2000qt,Contino:2006nn,Serone:2009kf}. Supersymmetric theory in very high energy/higher dimensions might be seen in an interesting and unusual way at the low energy, such as low energy theory being ``Emergent"\cite{Seiberg:2006wf,You:2014vea,Kaplan:2011vz}, different from the conventional compactification and supersymmetry breaking.  
 
 In reference \cite{2013PhRvD..87l5036S,Kaplan:2011vz} we took a modest low energy approach,  that proposed to extend the Little Higgs theory to  use the additional symmetries as flavor symmetries and related the quark masses to the structure of the soft symmetry breaking terms. We found that the resulting model had new flavor physics at the TeV scale but escaped low energy constraints on Flavor Changing Neutral Currents (FCNC).
 
\begin{figure}[t]
\includegraphics[width=5cm,natwidth=610,natheight=642]{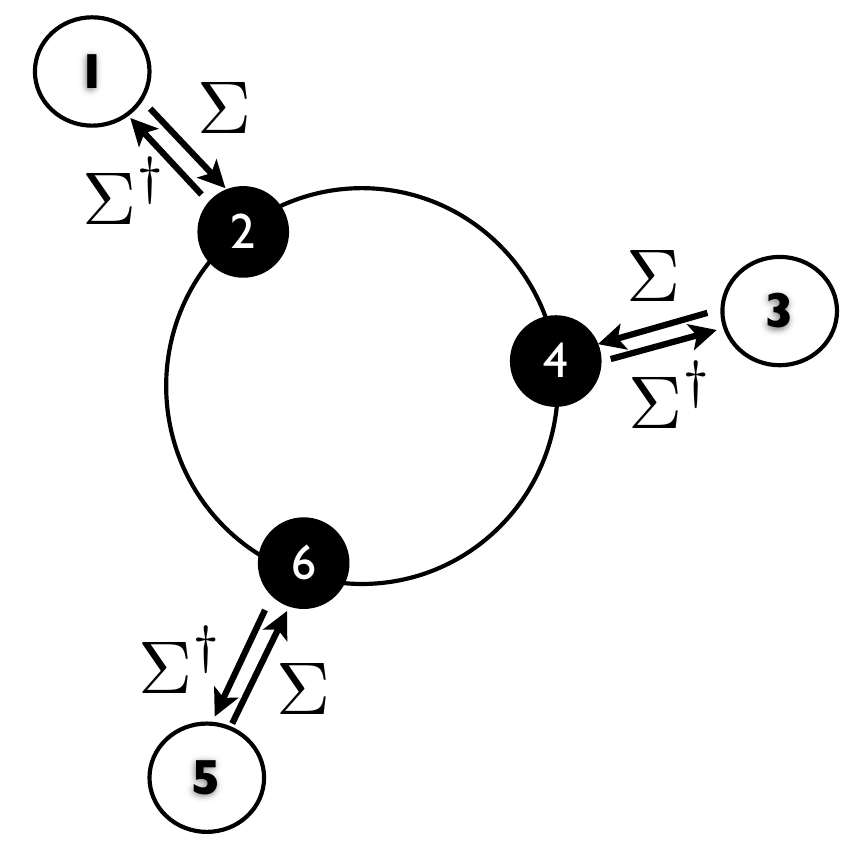}
\caption{ {A deformed moose diagram shows chiral fields on the white sites and their partners on the black sites. The big circle in the middle represents $SU(4)\times U(3)$ breaking terms }}
\label{fig:partner}
\end{figure}

  Fig.~\ref{fig:partner} shows a way to look at the Little Flavor model in terms of fermions, their partners and couplings. This figure resembles a moose/quiver graph with gauge groups living on each sites. The difference is that the sites are connected by non-linear $\sigma$ model fields which are in the bi-fundamental representations of the gauge groups. The fermions are in fundamental representations and live on the sites.
  
  Each generation of standard model fermions mostly live on a white site, with a seesaw type mixing with their own partners. The "partners" here mostly live on the adjacent black sites which are coupled to their own standard model companions on white sites through non-linear $\sigma$ model fields \textemdash the oriented lines in the graph. Each cell contains one black fermion and one white fermion. Due to the fermion multiplet structure, the small mixing across the cells is also directly proportional to the SU(4) symmetry breaking on black sites, which is proportional to the SM fermion mass. 
   
   The partner structure in Little Flavor is more complicated, because the breaking of the global symmetry associated with the fermion multiplets are not only responsible for generating the Higgs potential, but also give masses to the Standard Model fermions.   In the first Little Flavor model, the chiral fields on white sites have exactly the same field content as the Standard Model, with every one of those chiral fields having a Dirac partner on the black site.
   
  Since SM type mixing, e.g., CKM, PMNS matrix, coming from left-handed W current, are basically describing the misalignment between up-type and down-type fermions in the currents, it can still have considerable amount of flavor off-diagonal matrix elements, while the neutral current are largely suppressed by the small mass of fermions comparing to the cut-off scale.
  
\begin{figure}[t]
\includegraphics[width=7cm,natwidth=610,natheight=642]{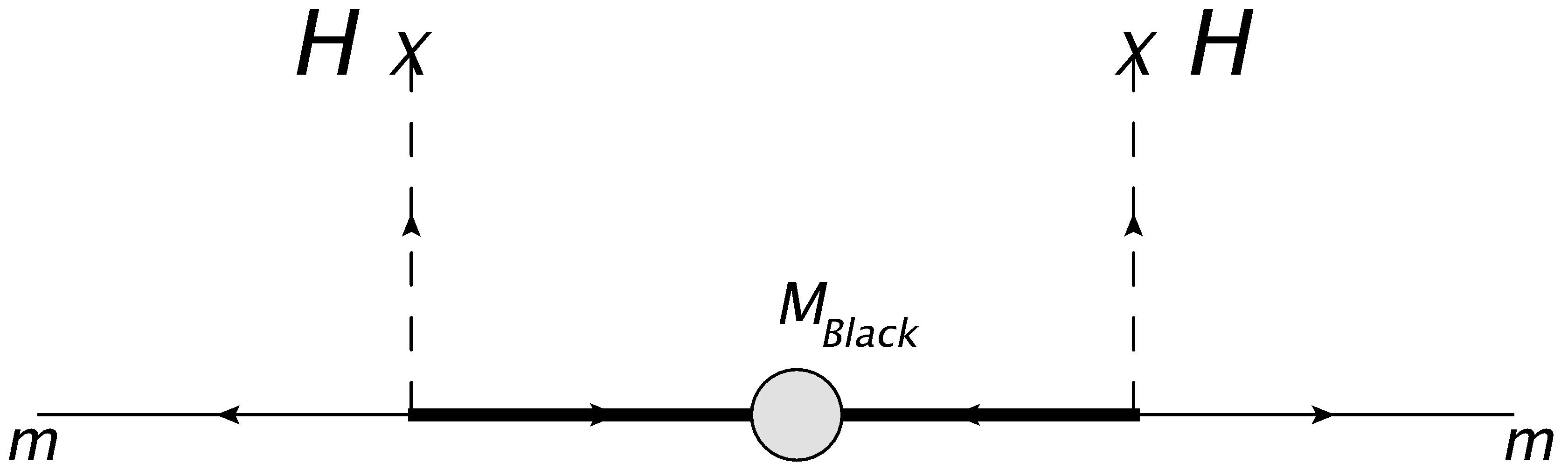}
\caption{Simplified effective couplings between the Standard Model fermions and their partners.}
\label{fig:eft}
\end{figure}

  Fig ~\ref{fig:eft} shows the Little Flavor model in an effective theory, which is further studied in \cite{Grabowska:2015rda}. The Standard Model fermions are coupled to Higgs fields and heavy partners with slightly broken global symmetries.
 
 In this paper, we extend the Little Flavor theory to include leptons, especially we explore a new type of neutrino seesaw by putting neutrinos in a SU(4) multiplet with charged leptons, identifying some fermion partners in the Little Flavor theory as right-handed Majorana neutrinos. The Little Flavor mechanism makes sure that the seesaw mass required by the neutrino mass bounds is no more than a couple of TeVs.
 We also discuss some phenomenological aspect of Little Flavor, including new search channels in $Z'$ physics, flavor violations and Higgs decay. Especially, the $Z'$ in the model could introduce a recent LHC anomaly in the angular distribution of B meson decay according to \cite{Descotes-Genon:2013wba,Buras:2013qja,Altmannshofer:2013foa,Gauld:2013qja}

\begin{figure}[t]
\includegraphics[width=6cm,natwidth=610,natheight=642]{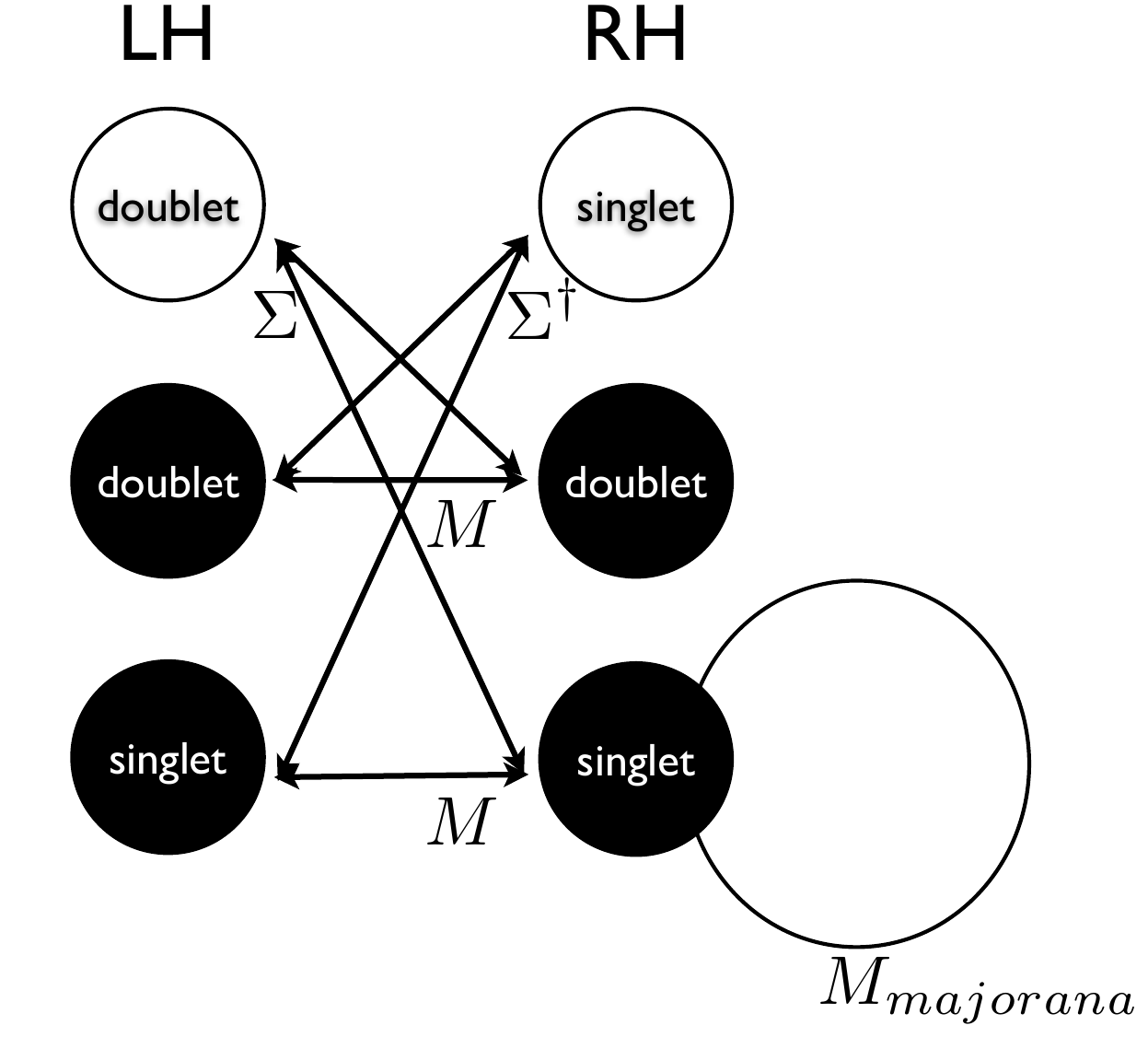}
\caption{ {A graph displays how upper or down type fermion fields respectively in one single cell are coupled to each others. All couplings are from the left-handed fermions to the right-handed fermions, except the right-handed neutrino Majorana mass term for black fermions, that gives neutrino mass extra suppression except for the black partner fermion "seesaw-like" mechanism.}}
\label{fig:coupling}
\end{figure}

\section{Gauge groups}

Here we work out the gauge sector in more details, to study the phenomenology. The couplings involving the  Higgs sector are discussed in \cite{Grabowska:2015rda}.  We further discuss an improved gauge $SU(2)\times SU(2)$ on black site at the end of gauge group discussion.

The gauge symmetry of the model is $G_w\times G_b$, where $G_{w,b}$ are independent $SU(2)\times U(1)$ groups living on white (w) and black (b) sites respectively. The $G_w\times G_b$ break down to the diagonal subgroup of the SM $SU(2)\times U(1)$, and  we take the the gauge couplings to be
\beq
g_{1,w}=\frac{g'}{\cos\gamma_1}\ ,\qquad g_{1,b} = \frac{g'}{\sin\gamma_1}\ ,\qquad 
g_{2,w}=\frac{g}{\cos\gamma_2}\ ,\qquad g_{2,b} = \frac{g}{\sin\gamma_2}\ .
\eqn{gamma}\eeq
 with $g=e/\sin\theta_w$ and $g'=e/\cos\theta_w$ are the usual SM gauge couplings and the angles $\gamma_{1,2}$ are free parameters. These couplings are designed to produce the right SM electroweak physics. 

The gauge fields are coupled to  an $SU(4)\times SU(4)/SU(4)$ nonlinear $\sigma$-model, parametrized by the field $\Sigma$, an $SU(4)$ matrix which transforms under $SU(4)\times SU(4)$ as the $(4,\mybar 4)$ representation, where Little Higgses live.  The $G_w\times G_b$ gauge symmetry is embedded in the $SU(4)\times SU(4)$ so that the covariant derivative acts on $\Sigma$ as
\beq
D_\mu\Sigma = \partial_\mu\Sigma +i\left(g_{2,w}A^a_\mu T_a + g_{1,w}  B_\mu Y\right) \Sigma - i\Sigma \left(g_{2,b}\tilde A^a_\mu T_a+g_{1,b} \tilde B_\mu Y\right)\ ,
\eeq
where $\{A^a_\mu, B_\mu\}$ and $\{\tilde A^a_\mu, \tilde B_\mu\}$ are the gauge bosons of $G_w$ and $G_b$ respectively.

The $\Sigma$ field spontaneously breaks $G_w\times G_b$ gauge symmetry down to a diagonal subgroup; if $\vev{\Sigma}=1$, the unbroken subgroup is the diagonal $SU(2)\times U(1)$, which is identified with the electroweak gauge group of the SM, and it has the correct couplings $g$ and $g'$. The small off-diagonal matrix elements of $\vev{\Sigma}=1$ further breaks EW $SU(2)\times U(1)$ to electromagnetic $U(1)$. There are two exotic $Z$ bosons and an exotic $W$ boson gotten masses from this spontaneously symmetry breaking, whose masses are:
\beq
M_{Z'} =\frac{g f}{\sin 2\gamma_2}\ ,\qquad
M_{Z''} = M_{W'}= \frac{g' f}{\sin 2\gamma_1}\ .
\eqn{heavyZ}\eeq
Electroweak symmetry breaking will correct these relations at $O(M^2_Z/f^2)$; in the model we consider in this paper we fix the Goldstone boson decay constant to be $f=1.5\TeV$; thus the corrections are $O(M^2_Z/f^2)\simeq 1\%$.

This is the same gauge sector we used in the previous quark model paper. The problem with this model is that the gauge couplings explicitly break black fermion $SU(4)$ at one-loop level. Because the key thing of Little Flavor mechanism is this black fermion $SU(4)$ protects fermion mass to prevent FCNC from generating, this is bad for the model and will radiatively generate fermion mass at hundreds of MeV.

One way to fix it is to gauge $SU(2)\times SU(2)$ at black site rather than $SU(2)\times U(1)$, and make upper $SU(2)$ coupling the same as lower $SU(2)$. Then there are two extra $W''$:
\beq
M_{W''}=M_{Z'} = \frac{g' f}{\sin 2\gamma_2}
\eeq
And let $g_{1,b} = g'/\sin\gamma_1=g_{2,b} = g/\sin\gamma_2$, which leads to:

\beq
\tan \theta_w= \frac{\sin \gamma_1}{\sin \gamma_2}
\eeq
This would not change phenomenological aspects much.

One could also work out the gauge boson self-coupling, here we stick to $SU(2)\times U(1)$ model:
\beq
&L \supset -\frac{1}{4} F^{\mu\nu}F_{\mu \nu}-\frac{1}{4} Z^{\mu\nu}Z_{\mu \nu}-\frac{1}{4} Z'^{\mu\nu}Z'_{\mu \nu}-\frac{1}{4}Z''^{\mu\nu}Z''_{\mu \nu}\\
& +\sum_{b,w}\{-D_i^{\dagger\mu}W_i^{-\nu}D_{\mu}W_i^{+\nu}+D_i^{\dagger\mu}W_i^{-\nu}D_{i \nu}W_i^{+\mu}+ig_{2i} F_{3i}^{\mu\nu} W_{i\mu}^{+} W^{-}_{i\nu} \}
\eeq
In the limit of $\gamma_1 \rightarrow \gamma_2 \rightarrow \gamma$ (one should note that if it is not in this limit, the mixings between $Z'Z''W'W''$ have non-trivial dependence on $\gamma_1$ and $\gamma_2$):
\beq
&W^\pm_w=-\sin\gamma W'^\pm+\cos\gamma W^\pm\\
&W^\pm_b=\cos\gamma W'^\pm+\sin \gamma W^\pm\\
&F_3^{\mu\nu}=\partial^\mu A_{3i}^\nu-\partial^\nu A_{3i}^\mu\\
&D_i=\partial_i-i g_i A_{3i}\\
&A_{3w}=\sin \gamma Z''+\cos \gamma \cos \theta_w Z+\cos \gamma \sin \theta_w A\\
&A_{3b}=-\cos \gamma Z''+\sin \gamma \cos \theta_w Z+\sin \gamma \sin \theta_w A
\eeq

The $\Sigma$ field describes fifteen pseudo Goldstone bosons with  decay constant $f$, to be set to $1.5\TeV$ in the phenomenological model we studied in the previous work  \cite{2013PhRvD..87l5036S}. With different parameter values $\gamma_1,\gamma_2$, the triple-gauge couplings present here can explain the recent diboson anomalies in $WW,ZZ,WZ$ channels \cite{Aad:2015owa,Hisano:2015gna,Dobrescu:2015qna,Brehmer:2015cia}. We are not exactly sure how the recent 750 GeV resonance \cite{atlas,CMS:2015dxe} fits into the model yet, possibly one of the Higgs partners in $\Sigma$ can be a good candidate. The parametrization of $\Sigma$ here is given by \cite{2013PhRvD..87l5036S}.

\begin{table}
\begin{tabular}{|c| c| c|c| } 
\hline
\quad& $WW$ & $WW'$& $W'W'$ \\
\toprule
$Z''$ & 0 & $-ie$ & $ie (\sin^3 \gamma/\cos \gamma-\cos^3 \gamma/\sin \gamma) $  \\ 
$Z'$ &0 &0 &0  \\
$Z$ &$ie\cot \theta_w$ & 0 & $ie \cot \theta_w$ \\
$A$ &$ie$ & 0 & $ie$ \\
\hline
 \end{tabular}
\caption{gauge bosons self-coupling in the limit $\gamma_1 \rightarrow \gamma_2 \rightarrow \gamma$. Notice here the physical gauge couplings also include the momentum factor $g_{\mu\nu}(k_2-k_1)_\rho+g_{\nu\rho}(k_3-k_2)_\mu+g_{\rho\mu}(k_1-k_3)_\nu$.} 
\label{tab:template} 
\end{table}

\section{fermion content and mass terms}
  
The model could be described by the deformed moose diagram of Fig.~\ref{fig:partner}, which is not a normally defined moose. The oriented links represent $\Sigma$ and $\Sigma^\dagger$, while the white and black sites represent fermions transforming nontrivially under $G_w$ and $G_b$ respectively. The big circle in the middle shows the mixing across the paired sites.

One could construct leptons as $SU(2)$ doublets $L = (n,e)$ and $SU(2)$ singlets $N,\,E$. The black fermions are Dirac fermions, which are made of left-handed and right-handed components, packed together as the quartets of approximate $U(4)_b$ symmetries
\beq
\psi_{b} = \begin{pmatrix} L\\ N\\ E \end{pmatrix}_b\ ,\qquad \text{black sites: }b=2,4,6\ .
\eqn{psi}\eeq

The white fermions are chiral fermions, which have the exactly same field content as the Standard Model fermions:

\beq
\chi_{w,L}  =  \begin{pmatrix} L\\0\\ 0 \end{pmatrix}_{w,L}\ ,\qquad 
\chi_{w,R}=  \begin{pmatrix} 0\\N\\ E \end{pmatrix}_{w,R}\ ,\qquad \text{white sites: }w=1,3,5\ .
\eqn{chi}\eeq

 $b=2,4,6$ and $w=1,3,5$ refer to the site numbers in Fig.~\ref{fig:partner}.

The $G_w$ and $G_b$ gauge generators are embedded within $U(4)_{w,b}$ as in $\Sigma$ field, except that $Y$  is extended to include a term   $\half(B-L)=\frac{1}{6}$ for all the fermions to produce the right hypercharge, this is universal for both quarks and leptons:
\beq
  Y =  \begin{pmatrix} 0 & 0\\0 & T_3\end{pmatrix} + \frac{1}{2} (B-L)=  \begin{pmatrix} 0 & 0\\0 & T_3\end{pmatrix} - \frac{1}{2} \begin{pmatrix} 1 & 0\\0 & 1\end{pmatrix}\qquad\text{(Leptons)}
\eqn{embed}
\eeq
where the leptonic fermions all carry $(B-L)=-1$. The $SU(2)$ part in $G_w \times G_b$ only gauges the upper part of the multiplets, the doublets.

To compact the mass term, we label the cells by $n=1,2,3$, with cell $n$ associated with sites $\{2n-1,2n\}$, and then an index $\alpha=1,2$ will specify the white and the black site respectively within the cell.  The fermions are all labeled then as $\Psi_{n,\alpha}$ with 
\beq
\Psi_{1,1} = \chi_{1} \ ,\quad
\Psi_{1,2} = \psi_{2} \ ,\quad
\Psi_{2,1} = \chi_{3} \ ,\quad
\Psi_{2,2} = \psi_{4} \ ,\quad
\Psi_{3,1} = \chi_{5} \ ,\quad
\Psi_{3,2} = \psi_{6} \ ,
\eeq
where the $\chi$ four-component  chiral fermions on the white sites in \eq{chi}, and the $\psi$ are the four component Dirac fermions on the black sites in \eq{psi}.

To get the electroweak symmetry breaking and neutrino Majorana masses we introduce two spurions to break the $SU(4)\times U(3)$ symmetry, defined by the traceless $4\times 4$ matrices which can be thought of as transforming as elements of the adjoint of $SU(4)$:
\beq
X_m=X_N = \begin{pmatrix} 
0& & &\cr
& 0 & &\cr
& & 1 &\cr
& & & 0
\end{pmatrix}\ ,\qquad
X_E = \begin{pmatrix} 
0 & & &\cr
& 0 & &\cr
& & 0 &\cr
& & & 1
\end{pmatrix}
\eeq
These matrices break the $SU(4)$ symmetry down to $SU(3)$ and will allow the light fermions to acquire masses; the $X_E$ matrix splits off the $E$ charged leptons from the $SU(4)$ multiplet, while $X_N$ distinguishes the $N$ neutrinos. We write all the mass term as below:

\beq
\CL=\CL_\text{sym}+ \CL_\text{asym}+ \CL_\text{asym,majorana} 
\eeq

The symmetric fermion mass and Yukawa terms are given by
\beq
\CL_\text{sym} =  \mybar \Psi_{m\alpha,L} \left[\CM_{m\alpha,n\beta} +\Sigma\, \CY_{m\alpha,n\beta} -\Sigma^\dagger\,\CY^\dagger_{m\alpha,n\beta}\right]  \Psi_{n\beta,R}\ ,
\eqn{Lsym}\eeq
where $\CM$, $\CY$ are independent and take the form
\beq
\CM_{m\alpha,n\beta}
= M_{black}  \,\begin{pmatrix} 1&&\cr &1&\cr&&1\end{pmatrix}_{mn}\otimes\ \ \begin{pmatrix}0 &0\\ 0& 1\end{pmatrix}_{\alpha\beta}
\eqn{cmval}\eeq
\beq
\CY =
 \lambda_{Yukawa}\, f\, \begin{pmatrix} 1&&\cr &1&\cr&&1\end{pmatrix}_{mn}\otimes\ \ \begin{pmatrix}0 &1\\ 0& 0\end{pmatrix}_{\alpha\beta} \ ,
\eqn{cyval}\eeq
where all unmarked matrix elements are zero.

The $\CM$ term is a common mass term for the black site Dirac fermions; the $\CY$ term is a nearest neighbor hopping interaction involving $\Sigma,\Sigma^\dagger$ in the direction of the link arrow. Note that these hopping terms combined look like a  covariant derivative in a fifth dimension, with $\Sigma$ playing the role of the fifth component of a gauge field. This could also be interpreted as the terms in lattice gauge theory with finite lattice sites.

 $X_m$ projects out neutrino field. We take for our symmetry breaking mass terms
\beq
\CL_\text{asym} =\mybar \Psi_{m\alpha,L} \left[ \CM^{E}_{m\alpha,n\beta}+ \CM^{N}_{m\alpha,n\beta} \right]\Psi_{n\beta,R}  
\eqn{Lasym}\eeq
where
\beq
\CM^{E}_{m\alpha,n\beta} = M^E_{mn}
\otimes \begin{pmatrix}0 & 0\cr 0 & 1 \end{pmatrix}_{\alpha\beta}\otimes X_E \ ,\qquad
\CM^{N}_{m\alpha,n\beta} = M^N_{mn}
\otimes \begin{pmatrix}0 & 0\cr 0 & 1 \end{pmatrix}_{\alpha\beta}\otimes X_N \ .
\eeq
Those terms are the couplings between black fermions. It is represented in the big circle in Fig.~\ref{fig:partner}. The  $\CM^{E}_{m\alpha,n\beta},\CM^{N}_{m\alpha,n\beta}$ are $3\times3$ matrices which contain textures.

Extra symmetry breaking majorana mass term:
\beq
\CL_\text{asym,majorana} =\Psi^T_{m\alpha,R} ( \CM^{m}_{m\alpha,n\beta})\Psi_{n\beta,R} 
\eqn{Lasym}\eeq
\beq
\CM^{m}_{m\alpha,n\beta} = M^m_{mn}
\otimes \begin{pmatrix}0 & 0\cr 0 & 1 \end{pmatrix}_{\alpha\beta}\otimes X_{m} \ .
\eeq

\beq
M^m_{mn}
= M_{majorana}  \,\begin{pmatrix} 1&&\cr &1&\cr&&1\end{pmatrix}_{mn}
\eqn{cmval}\eeq

Where $X_{m}$ projects out neutrino field $N$ of whole multiplet, and $\CM^{M}_{m\alpha,n\beta}$ is a real Majorana mass matrix. 

By having the $X$ spurions each leave intact an $SU(3)$ subgroup of the black-site $SU(4)$ symmetry, we ensure that the fermions will not contribute any one-loop quadratically divergent mass contributions to the Higgs boson (the Little Higgs mechanism). Although it is not important in the lepton sector at current TeV scale due to the smallness of lepton masses. The extra ingredient here comparing to the quark sector is the Majorana masses. We will show how to construct the mass matrix  from this Lagrangian later in the Appendix.

\section{A texture and effective mass formulas }

We can choose a simplest texture which agrees with experimental results on lepton masses and mixing matrix. This choice is not unique. The interesting thing about this particular texture is to explicitly put all the flavor violation effect in neutrino sector 
to evade the stringent bounds in charged lepton sector:
\beq
M^{E} = D_E,\qquad
M^{N} = U_{NH}^T \cdot D_N\ .
\eeq
With
\beq
D_E=\begin{pmatrix} M_e&&\cr &M_{\mu}&\cr&&M_{\tau}\end{pmatrix}, \quad
D_N=\begin{pmatrix} M_{\nu_e}&&\cr &M_{\nu_\mu}&\cr&&M_{\nu_\tau}\end{pmatrix},\quad
\eeq
And $U_{NH}$ being the unitary lepton mixing matrix for normal hierarchy. One can also chose the texture for the inverse hierarchy. 

In this model, one can easily integrating out heavy particles, and all the Standard Model lepton masses are perturbative. Following the analysis of EFT , we arrive at mass
formulas for neutrinos and leptons:
\beq
& m_n=\frac{(D_N)^2}{M_{majorana}}\frac{v_n^2}{f^2}\frac{(\lambda_{Y} f)^2}{(\lambda_{Y} f)^2+M_{black}^2}[1+O(\frac{M^2_{black}}{M^2_{majorana}})]\\
& m_e=\frac{v_n}{f}\frac{(\lambda_{Y} f)^2}{(\lambda_{Y} f)^2+M_{black}^2} D_E [1+O(\frac{D_E}{M_{black}})]
\eeq

 Note see-saw mechanism from right-handed $M^m$ for neutrinos, and the extra factor $(\lambda f)^2/[(\lambda f)^2+M^2]$ from wave 
function renormalization, equivalent to the effect of mixing with heavy vector-like leptons. It is also straightforward to see that this texture correctly produces the 
neutrino mixing matrix $U_{NH}$.

\section{A realistic solution}

In order to show the spectrum of all the new particles and compute heavy gauge boson couplings to the fermions, we choose to parametrize the model as below

\beq
f=1500\text{ GeV},\quad \lambda_{Y,Lepton}= 0.1498\\
M_{black}= 500\text{ GeV},\quad tan\beta=\frac{v_n}{v_e}=1\\
 M_{majorana}=5000\text{ GeV}
\eeq
Note that the size of composite coupling $f$ is shared with quark sector, while leptons have their own universal Yukawa coupling, which is one-tenth of quark Yukawa. The mass scale of heavy lepton partners is $500$ GeV, which makes the lightest lepton partners be about $200$ GeV. Considering the experimental bound of charged leptons, about $100$ GeV \cite{Nakamura:2010zzi}, It is also possible in the model to have even lighter lepton partners. 

To have the lighter fermion partners, one could tune down $M_{black}$, then it is necessary to also have smaller $\lambda_{Y,Lepton}$ due to the fact that the ratio $M_{black}/\lambda f$ is directly related to the mixing of right-handed and left-handed particles. If the SM left-handed particles have too much right-handed component which is coupled to W boson, we will end up with non-unitary mixing matrix.  

Choose the input in heavy lepton mass matrix as the PDG valued $U_{NH}$
\beq
U_{NH}=
\left(
\begin{array}{ccc}
 0.822 & 0.547 & -0.15 \\
 -0.355 & 0.702 & 0.616 \\
 0.442 & -0.452 & 0.772 \\
\end{array}
\right)
\eeq
Then one will have SM lepton mixing matrix as:
\beq
U'_{NH}=\left(
\begin{array}{ccc}
 0.818 & 0.551 & -0.147 \\
 -0.363 & 0.701 & 0.61 \\
 0.44 & -0.446 & 0.776 \\
\end{array}
\right)
\eeq
$U'_{NH}$ is the mixing matrix computed from the full theory. In Appendix.A we show the detail of computation: generating the mixing matrix and mass spectrum from full theory. Notice that
\beq
U'_{NH} U'^{\dagger}_{NH}=
\left(
\begin{array}{ccc}
 0.9950 & 0 & 0 \\
 0 &0.9950 & 0 \\
 0 & 0 & 0.9957 \\
\end{array}
\right)
\eeq
The non-unitarity of PMNS matrix at $0.5 \%$  is due to both the mixing with heavy lepton partners and right-handed particles. The non-unitarity bound for the PMNS matrix is discussed in \cite {Antusch:2014woa,Antusch:2006vwa}. Although $0.5 \%$ seems to be in tension with the newest bounds, we conclude it is still not clear about the unitarity bound in the LF model, because  \cite {Antusch:2014woa,Antusch:2006vwa} assumed only the SM particle spectrum.

We also choose the eigenvalues of the mass texture to be as below, these are the parameters that feed into the SM particle masses.

\beq
& D_N=\left(
\begin{array}{ccc}
 2.278\times 10^1 & 0 & 0 \\
 0 & 2.319\times 10^1 & 0 \\
 0 & 0 & 2.541\times 10^1 \\
\end{array}
\right)\text{ MeV},\\
& D_E=\left(
\begin{array}{ccc}
 2.611\times 10^1 & 0 & 0 \\
 0 & 5.444\times 10^3 & 0 \\
 0 & 0 & 1.069\times 10^5 \\
\end{array}
\right)\text{ MeV}
\eeq

The full mass spectrum, including all the partners are, For Dirac-like charged leptons (in GeV):
\beq
&(647.2,553.1,548.2,548.2,548.2,548.2,1.773,1.056\times 10^{-1},5.110\times 10^{-4})
   \eeq
 For neutrinos, note that there is the mass splitting for right-handed and left-handed particles:
 \beq
 \nonumber  &(5050,5050,5050,548.2,548.2,548.2,\\
 \nonumber &548.2,548.2,548.2,
    248.3,248.3,248.3,\\
 &198.6,198.6,198.6,2.974\times 10^{-10},2.482\times 10^{-10},2.393\times
   10^{-10})\text{ GeV} \label{spectrum}
\eeq

\section{$Z'$,$Z''$ and $W'$ }

Here we choose $\gamma_1=\gamma_2=\pi/8$, to be comparable to the quark sector, which gives rise to $M_{Z'} = 750\GeV, M_{Z''} = 1400\GeV$. One could easily go to different values of $\gamma_1, \gamma_2$ to accommodate $SU(2)\times SU(2)$ model. The gauge bosons are coupled to the Standard Model fermions as below, as the vector basis being $\{e,\mu,\tau\}$ and $\{\nu_e, \nu_\mu, \nu_\tau\}$. The neutral gauge boson couplings are strictly flavor diagonal, due to to the choice of $M^N$ in this paper, which only has the texture matrix on the left. 

Another interesting thing is the reduced $Z',Z''$ coupling comparing to the SM Z, which is due to the choice of free parameters $\gamma_1, \gamma_2$, one can see the further analysis in \cite{Grabowska:2015rda}. One could fine-tune them to completely decouple $Z',Z''$, but it is not necessary. We can see later that even SM model-like couplings are enough to evade the current bounds because of the extensive lepton sector. 

\begin{align}
 &L^n_{Z}=\left(
\begin{array}{ccc}
 3.627 & 0 & 0 \\
 0 & 3.627& 0 \\
 0 & 0 & 3.627\\
\end{array}
\right)\times 10^{-1} \\
&L^n_{Z''}=\left(
\begin{array}{ccc}
 -1.919 & 0 & 0 \\
 0 & -1.919 & 0 \\
 0 & 0 & -1.919 \\
\end{array}
\right)\times 10^{-2} , 
L^n_{Z'}=\left(
\begin{array}{ccc}
 1.229 & 0 & 0 \\
 0 & 1.229 & 0 \\
 0 & 0 & 1.229 \\
\end{array}
\right)\times 10^{-2}
\end{align}

\begin{align}
R^e_{Z}=\left(
\begin{array}{ccc}
 1.613 & 0 & 0 \\
 0 & 1.613 & 0 \\
 0 & 0 & 1.617 \\
\end{array}
\right)\times 10^{-1}, 
L^e_{Z}=\left(
\begin{array}{ccc}
 -2.006 & 0 & 0 \\
 0 & -2.006 & 0 \\
 0 & 0 & -2.008 \\
\end{array}
\right)\times 10^{-1}
\end{align}

\begin{align}
R^e_{Z''}=\left(
\begin{array}{ccc}
 1.204 & 0 & 0 \\
 0 & 1.197 & 0 \\
 0 & 0 & 1.078 \\
\end{array}
\right)\times 10^{-3}, 
L^e_{Z''}=\left(
\begin{array}{ccc}
 1.86 & 0 & 0 \\
 0 & 1.859 & 0 \\
 0 & 0 & 1.851 \\
\end{array}
\right)\times 10^{-2},
\end{align}

\begin{align}
R^e_{Z'}=\left(
\begin{array}{ccc}
 2.12 & 0 & 0 \\
 0 & 1.834 & 0 \\
 0 & 0 & -2.43 \\
\end{array}
\right)\times 10^{-2},
L^e_{Z'}=\left(
\begin{array}{ccc}
 9.896& 0 & 0 \\
 0 & 9.859 & 0 \\
 0 & 0 & 9.296 \\
\end{array}
\right)\times 10^{-3}
\end{align}

We can define the charged current according to the normalization below:
\beq
-\frac{g_2}{\sqrt{2}} \,\left({W}^+_\mu \bar N_i \CL^{W}_{ij} \gamma^\mu P_L E_j + {W'}^+_\mu \bar N_i \CL^{W'}_{ij} \gamma^\mu P_L E_j\right).
\eeq
\beq
L_{W}=\left(
\begin{array}{ccc}
 7.902\times 10^{-1} & 5.324\times 10^{-1} & -1.422\times 10^{-1} \\
 -3.504\times 10^{-1} & 6.773\times 10^{-1} & 5.889\times 10^{-1}
   \\
 4.253\times 10^{-1} & -4.312\times 10^{-1} & 7.494\times 10^{-1} \\
\end{array}
\right), \\
L_{W'}=\left(
\begin{array}{ccc}
 4.652\times 10^{-2} & 3.134\times 10^{-2} & -8.35\times 10^{-3} \\
 -2.063\times 10^{-2} & 3.987\times 10^{-2} & 3.457\times 10^{-2}
   \\
 2.504\times 10^{-2} & -2.538\times 10^{-2} & 4.4\times 10^{-2} \\
\end{array}
\right) \approx L_W \frac{M_W}{M_{W'}}
\eeq
Notice that $L_W \approx V_{pmns}$. There are couples of interesting physics process related to these couplings:

\begin{enumerate}
\item
As in previously discussed quark sector, $Z'$ couplings to quarks are smaller than $Z$ couplings by a factor of around $10^{-2}$, and the $Z''$ and $W'$ couplings to quarks are suppressed by about an order of magnitude relative to the $Z$ and $W$ couplings. These suppressions reduce the production rates of new gauge boson dramatically.  They are also enough to satisfy current collider bounds for jet, top quark and gauge boson final states \cite{Chatrchyan:2012rva,CMS:2012xva,Aad:2012raa,Aad:2013wxa,Aad:2013nca,Chatrchyan:2013qha}. \\ 
\item
 We can see here $Z',Z'', W'$ couplings to leptons are suppressed by at least an order of magnitude relative to the $Z$ and $W$ couplings. More importantly, the branching ratios to the Standard model leptons are quite reduced, because all the lepton partners except RH Majorana neutrinos in current parameter choices are lighter than these new gauge bosons, which could have more decay channels rather than those in the Standard model.Thus those $Z',Z''$ are still allowed for the current dilepton final states of $Z'$ search  \cite{Chatrchyan:2012oaa,Aad:2012hf,CMS-PAS-EXO-12-061,2012PhLB..714..158C,atlas13}, including the newest 13TeV run II \cite{atlas13}. Moreover, the reduced coupling of $Z'$ to leptons will also satisfy the electroweak precision constraints. See the discussion of constraints on $Z'$ in \cite{Langacker:2008yv}.
 \item
We can also see that the new physics mediated FCNC are at loop level through $W'$ exchange, or with heavy leptons mediated penguin diagrams. Note that there is similar GIM suppression with $L_{W'}$, new physics FCNC contribution will occur at the same loop level but much smaller than the standard model ones.
\item
It is not likely to see tree level flavor-changing charged current contribution to neutrino oscillation experiments in the current model set-up due to $L_{W'} \varpropto L_W$. With more texture built into $M^E, M^N$, we will be able to see the the deviation from long base line neutrino oscillation experiments due to $W'$ mediated oscillations.
\item
There are non-zero coupling between the $Z',Z'',W'$ and heavy lepton partners, both diagonal and off-diagonal ones. One could calculate it to estimate the size of different branching ratio for other channels of new gauge bosons. We will talk about it more in the next section.
\item
 We expect sizable contributions to the muon magnetic moment rising from the $Z'$ gauge boson, in particular. Although the $M_Z'$ is at the TeV scale, those corrections are actually proportional to $M^2_{L}/M_{Z'}^2$, which can be big enough to explain the muon magnetic moment anomaly. This discussion is further explored in \cite{Allanach:2015gkd}.
\end{enumerate}

\section{New physics process}

\begin{table}
\begin{tabular}{|c| c| c|c|c|c|c|c|c|c|} 
\hline
\quad&\multicolumn{2}{c|}{$ll$ }&\multicolumn{2}{c|}{$Ll$ }& \multicolumn{2}{c|}{$LL$ } 
&\multicolumn{1}{c|}{$NN$ }&\multicolumn{1}{c|}{$Nn$ } &\multicolumn{1}{c|}{$nn$ }  \\
\hline
\quad&L&R&L&R &L&R&&&\\
\hline
\hline
$Z$ &0.20&0.161&$=<0.02$ &$=<0.02$ &0.2&0.2    &0.36 &$=<0.01$&0.36 \\ 
$Z''$ &0.019&0.0011&$=<0.34$&$=<0.04$&0.63  &0.77  &0.70&$=<0.02$&0.019 \\ 
$Z'$ &0.0098 &0.02 &$=<0.18$&$=<0.4$&0.83 &0.72  &0.378&$=<0.13$&0.012  \\
\hline
 \end{tabular}
\caption{$Z',Z''$ couplings at tree level} 
\label{tab:template} 
\end{table}

\begin{table}
\begin{tabular}{|c| c| c|c|c| } 
\hline
\quad& $ln$ & $Ln$& $lN$ & $LN$ \\
\hline
$W$ & $V_{pmns}$ &$\sim V_{pmns} M_w/M_{w'}$ &$=<0.02$ & $=<0.6$  \\ 
$W'$ &$V_{pmns} M_w/M_{w'}$ &$\sim V_{pmns}$ &$\sim V_{pmns}$& $=<1$  \\
\hline
\end{tabular}
\caption{$W'$ exotic decay channels} 
\label{tab:template} 
\end{table}

\begin{table}
\begin{tabular}{|c| c| c|c|c|c| } 
\hline
\quad&\multicolumn{1}{c|}{$ll$ }&\multicolumn{2}{c|}{$Ll$ }& \multicolumn{1}{c|}{$LL$}\\
\hline
$h_{sm}$ & $\lambda_{yukawa}$ & $\sin \frac{v}{f}$ & $\cos \frac{v}{f}$ & $=<0.01$ \\
\hline
\end{tabular}
\caption{ Higgs couplings. Note that $\lambda_{yukawa} \sim 0.01$ for $\tau$ in the Standard Model. All the couplings are flavor-diagonal. The heavy to light couplings are only between SM fermions and their own partners.} 
\label{tab:template} 
\end{table}

A couple of hundreds GeV lepton partners can mediate some new processes. Here below $L$ or $N$ stands for heavy exotic charged lepton partners and heavy neutrinos while $l,n$ are the SM leptons. For $Z'$ search, the new processes could be summarized as below: 
\begin{enumerate}
\item
$Z' \rightarrow Ll \rightarrow nnll $ or $Z'' \rightarrow W'^+W'^- \rightarrow nnll $ \\
 \item
$Z' \rightarrow NN \rightarrow nnllll $ \\
It is also possible to have longer decay chain of $Z'$, to have more than 4 leptons final states.
\item
$Z'' \rightarrow W^+W'^- \rightarrow W^+lN\rightarrow W^+W^+ll\rightarrow 4jets+ll $ \\
\end{enumerate}

For $W'$ search, there are also multi-lepton final states:  $W'\rightarrow Nl \rightarrow nlll $ or $W'\rightarrow nL \rightarrow nnnl $.

From the tree level couplings listed in the table, one could see that over $95\%$ $Z',Z'',W'$ decays go to multi-leptons and $E_{tmiss}$ final states rather than the Standard Model type di-muons or $e^+e^-$, which are under $1\%$. Detailed branching ratios depend on specific parameters in the model, such as $\gamma_1, \gamma_2$ in the gauge sector. Multi-leptons plus missing energy final states have been heavily studied in supersymmetry search \cite{Aad:2014qaa,Aad:2014vma,Aad:2014iza}, and some recent study in the context of vector-like quark extension \cite{Chen:2015jmn,Berger:2014gga}. $Z' Z'' W'$ masses being 700 and 1400 $GeV$ are above the current bounds, but it may provide future guide for $Z'$ search. Especially reconstructing two same sign $W$ from $ 4jets+ll$ final states could be a distinct signature for this type of models.

Although it is not in this particular choice of parameters, but it is also possible to have lepton partners even lighter than the Higgs boson, but heavier than current bounds from LEP \cite{Nakamura:2010zzi}. The $L(^*)$ or $N(^*)$ stands for a heavy lepton or a virtual heavy lepton.
\begin{enumerate}
\item
$h \rightarrow L (^*)l, N(^*)n \rightarrow nnll \rightarrow \slashed E_T +ll$\\
\end{enumerate}

 This type of exotic decay final states of Higgs exists in some supersymmetry models as discussed in \cite{Curtin:2013fra,Huang:2013ima}. However, due to the large $Llh$ couplings as in the table below, this scenario is most likely to be ruled out for heavy leptons mass smaller than 125 $GeV$ from current Higgs branching ratio data. The lepton partners behave much like top/quark-partners in terms of the Higgs couplings. The vector-like quarks are constraint by LHC \cite{Moreau:2012da,Angelescu:2015kga}, and there are extensive literatures about top-partner search e.g.\cite{Anandakrishnan:2015yfa,Berger:2012ec}. The lepton partners are less discussed, and will be more probed in future generation colliders like ILC, CEPC and FCC-ee.

\section{conclusion}
The Little Flavor model \cite{Kaplan:2011vz,2013PhRvD..87l5036S} provides a new direction for the TeV flavor physics, while in the same time extends the Little Higgs models. It also alleviates the fine-tuning problem in the Little Higgs theory.  Fermions get masses in this new mechanism through the breaking of vector-like symmetries associated with exotic vector-like Dirac fermions and an approximate nonlinear symmetry in the Higgs sector. It suppresses FCNC only by a $U(2)^2$ vector symmetry rather than $U(2)^5$ chiral symmetry in Minimal Flavor Violation in the low energy \cite{Grabowska:2015rda}, it also provides multiple collider signatures within LHC reach. 

In this paper, we focus on the lepton part of the flavor model, and constructed a TeV neutrino see-saw mechanism. The new physics beyond the SM could be observed through $Z'$ and Higgs physics, as experimental proofs. 
 
 \begin{acknowledgments}
 I would like to thank David B. Kaplan, Ann E. Nelson, Martin Schmaltz and Tao Liu for useful discussion. I especially thank Martin Schmaltz for suggesting Figure 1. This work was supported in part by U.S. DOE Grant No. DE-FG02-00ER41132 and by the CRF Grants of the Government of the Hong Kong SAR under HUKST4/CRF/13G. 
 \end{acknowledgments}

 \appendix
 \section{Generating 18 $\times$ 18 Mass matrix with Majorana masses}
 
 Fig ~\ref{fig:coupling} shows that the couplings between fermions we wrote down. Except the Majorana mass, all the couplings are between left-handed and right-handed particles. For neutrinos, there are totally 9 left-handed field, 3 on each paired site, and 9 right-handed ones. One can  construct the 18-vector, with 9 left-handed fields first and then 9 right-handed fields:
 \beq
 F=\{L^L_{n,w},L^L_{n,b}, N^L_{b},......,N^R_{w},L^R_{n,b}, N^R_{b},......\}
 \eeq
 
 Then the mass matrix can be written in a compact form:
 \beq
 \CL =F^T M_{18\times18} F +h.c.
 \eeq
 Diagonalizing the mass matrix yields the neutrino spectrum in eq. \ref{spectrum}. According to the parameter choice we used in this paper, the numbers for the mass matrices are as below (in unit GeV):
 
 \beq
 M_1=\left(
\begin{smallmatrix}
 0. & 223.149 & 26.3591 & 0. & 0. & 0. & 0. & 0. & 0. \\
 26.3591 & 500. & 0. & 0. & 0. & 0. & 0. & 0. & 0. \\
 -223.149 & 0. & 500.019 & 0. & 0. & -0.00823179 & 0. & 0. & 0.0112327 \\
 0. & 0. & 0. & 0. & 223.149 & 26.3591 & 0. & 0. & 0. \\
 0. & 0. & 0. & 26.3591 & 500. & 0. & 0. & 0. & 0. \\
 0. & 0. & 0.0124598 & -223.149 & 0. & 500.016 & 0. & 0. & -0.0114868 \\
 0. & 0. & 0. & 0. & 0. & 0. & 0. & 223.149 & 26.3591 \\
 0. & 0. & 0. & 0. & 0. & 0. & 26.3591 & 500. & 0. \\
 0. & 0. & -0.00341675 & 0. & 0. & 0.0142839 & -223.149 & 0. & 500.02 \\
\end{smallmatrix} 
\right)
\eeq
 
 \beq
M_2= \left(
\begin{smallmatrix}
 0 & 0 & 0 & 0 & 0 & 0 & 0 & 0 & 0 \\
 0 & 0 & 0 & 0 & 0 & 0 & 0 & 0 & 0 \\
 0 & 0 & 5000. & 0 & 0 & 0 & 0 & 0 & 0 \\
 0 & 0 & 0 & 0 & 0 & 0 & 0 & 0 & 0 \\
 0 & 0 & 0 & 0 & 0 & 0 & 0 & 0 & 0 \\
 0 & 0 & 0 & 0 & 0 & 5000. & 0 & 0 & 0 \\
 0 & 0 & 0 & 0 & 0 & 0 & 0 & 0 & 0 \\
 0 & 0 & 0 & 0 & 0 & 0 & 0 & 0 & 0 \\
 0 & 0 & 0 & 0 & 0 & 0 & 0 & 0 & 5000. \\
\end{smallmatrix} 
\right) 
\eeq

\beq
M_{18\times 18}=\left(
\begin{array}{cc}
 0 & M_1 \\
  M^\dagger_1& M_2\\
   \end{array}
\right)
\eeq

\bibliography{flavorbib}

\begin{thebibliography}{87}%
\makeatletter
\providecommand \@ifxundefined [1]{%
 \@ifx{#1\undefined}
}%
\providecommand \@ifnum [1]{%
 \ifnum #1\expandafter \@firstoftwo
 \else \expandafter \@secondoftwo
 \fi
}%
\providecommand \@ifx [1]{%
 \ifx #1\expandafter \@firstoftwo
 \else \expandafter \@secondoftwo
 \fi
}%
\providecommand \natexlab [1]{#1}%
\providecommand \enquote  [1]{``#1''}%
\providecommand \bibnamefont  [1]{#1}%
\providecommand \bibfnamefont [1]{#1}%
\providecommand \citenamefont [1]{#1}%
\providecommand \href@noop [0]{\@secondoftwo}%
\providecommand \href [0]{\begingroup \@sanitize@url \@href}%
\providecommand \@href[1]{\@@startlink{#1}\@@href}%
\providecommand \@@href[1]{\endgroup#1\@@endlink}%
\providecommand \@sanitize@url [0]{\catcode `\\12\catcode `\$12\catcode
  `\&12\catcode `\#12\catcode `\^12\catcode `\_12\catcode `\%12\relax}%
\providecommand \@@startlink[1]{}%
\providecommand \@@endlink[0]{}%
\providecommand \url  [0]{\begingroup\@sanitize@url \@url }%
\providecommand \@url [1]{\endgroup\@href {#1}{\urlprefix }}%
\providecommand \urlprefix  [0]{URL }%
\providecommand \Eprint [0]{\href }%
\providecommand \doibase [0]{http://dx.doi.org/}%
\providecommand \selectlanguage [0]{\@gobble}%
\providecommand \bibinfo  [0]{\@secondoftwo}%
\providecommand \bibfield  [0]{\@secondoftwo}%
\providecommand \translation [1]{[#1]}%
\providecommand \BibitemOpen [0]{}%
\providecommand \bibitemStop [0]{}%
\providecommand \bibitemNoStop [0]{.\EOS\space}%
\providecommand \EOS [0]{\spacefactor3000\relax}%
\providecommand \BibitemShut  [1]{\csname bibitem#1\endcsname}%
\let\auto@bib@innerbib\@empty
\bibitem [{\citenamefont {{Sun}}\ \emph {et~al.}(2013)\citenamefont {{Sun}},
  \citenamefont {{Kaplan}},\ and\ \citenamefont
  {{Nelson}}}]{2013PhRvD..87l5036S}%
  \BibitemOpen
  \bibfield  {author} {\bibinfo {author} {\bibfnamefont {S.}~\bibnamefont
  {{Sun}}}, \bibinfo {author} {\bibfnamefont {D.~B.}\ \bibnamefont {{Kaplan}}},
  \ and\ \bibinfo {author} {\bibfnamefont {A.~E.}\ \bibnamefont {{Nelson}}},\
  }\href {\doibase 10.1103/PhysRevD.87.125036} {\bibfield  {journal} {\bibinfo
  {journal} {\prd}\ }\textbf {\bibinfo {volume} {87}},\ \bibinfo {eid} {125036}
  (\bibinfo {year} {2013})},\ \Eprint {http://arxiv.org/abs/1303.1811}
  {arXiv:1303.1811 [hep-ph]} \BibitemShut {NoStop}%
\bibitem [{\citenamefont {Kaplan}\ and\ \citenamefont
  {Georgi}(1984)}]{Kaplan:1983fs}%
  \BibitemOpen
  \bibfield  {author} {\bibinfo {author} {\bibfnamefont {D.~B.}\ \bibnamefont
  {Kaplan}}\ and\ \bibinfo {author} {\bibfnamefont {H.}~\bibnamefont
  {Georgi}},\ }\href {\doibase 10.1016/0370-2693(84)91177-8} {\bibfield
  {journal} {\bibinfo  {journal} {Phys.Lett.}\ }\textbf {\bibinfo {volume}
  {B136}},\ \bibinfo {pages} {183} (\bibinfo {year} {1984})}\BibitemShut
  {NoStop}%
\bibitem [{\citenamefont {Kaplan}\ \emph {et~al.}(1984)\citenamefont {Kaplan},
  \citenamefont {Georgi},\ and\ \citenamefont {Dimopoulos}}]{Kaplan:1983sm}%
  \BibitemOpen
  \bibfield  {author} {\bibinfo {author} {\bibfnamefont {D.~B.}\ \bibnamefont
  {Kaplan}}, \bibinfo {author} {\bibfnamefont {H.}~\bibnamefont {Georgi}}, \
  and\ \bibinfo {author} {\bibfnamefont {S.}~\bibnamefont {Dimopoulos}},\
  }\href {\doibase 10.1016/0370-2693(84)91178-X} {\bibfield  {journal}
  {\bibinfo  {journal} {Phys.Lett.}\ }\textbf {\bibinfo {volume} {B136}},\
  \bibinfo {pages} {187} (\bibinfo {year} {1984})}\BibitemShut {NoStop}%
\bibitem [{\citenamefont {Arkani-Hamed}\ \emph
  {et~al.}(2002{\natexlab{a}})\citenamefont {Arkani-Hamed}, \citenamefont
  {Cohen}, \citenamefont {Katz},\ and\ \citenamefont
  {Nelson}}]{ArkaniHamed:2002qy}%
  \BibitemOpen
  \bibfield  {author} {\bibinfo {author} {\bibfnamefont {N.}~\bibnamefont
  {Arkani-Hamed}}, \bibinfo {author} {\bibfnamefont {A.}~\bibnamefont {Cohen}},
  \bibinfo {author} {\bibfnamefont {E.}~\bibnamefont {Katz}}, \ and\ \bibinfo
  {author} {\bibfnamefont {A.}~\bibnamefont {Nelson}},\ }\href@noop {}
  {\bibfield  {journal} {\bibinfo  {journal} {JHEP}\ }\textbf {\bibinfo
  {volume} {0207}},\ \bibinfo {pages} {034} (\bibinfo {year}
  {2002}{\natexlab{a}})},\ \Eprint {http://arxiv.org/abs/hep-ph/0206021}
  {arXiv:hep-ph/0206021 [hep-ph]} \BibitemShut {NoStop}%
\bibitem [{\citenamefont {Arkani-Hamed}\ \emph
  {et~al.}(2002{\natexlab{b}})\citenamefont {Arkani-Hamed}, \citenamefont
  {Cohen}, \citenamefont {Katz}, \citenamefont {Nelson}, \citenamefont
  {Gregoire} \emph {et~al.}}]{ArkaniHamed:2002qx}%
  \BibitemOpen
  \bibfield  {author} {\bibinfo {author} {\bibfnamefont {N.}~\bibnamefont
  {Arkani-Hamed}}, \bibinfo {author} {\bibfnamefont {A.}~\bibnamefont {Cohen}},
  \bibinfo {author} {\bibfnamefont {E.}~\bibnamefont {Katz}}, \bibinfo {author}
  {\bibfnamefont {A.}~\bibnamefont {Nelson}}, \bibinfo {author} {\bibfnamefont
  {T.}~\bibnamefont {Gregoire}},  \emph {et~al.},\ }\href@noop {} {\bibfield
  {journal} {\bibinfo  {journal} {JHEP}\ }\textbf {\bibinfo {volume} {0208}},\
  \bibinfo {pages} {021} (\bibinfo {year} {2002}{\natexlab{b}})},\ \Eprint
  {http://arxiv.org/abs/hep-ph/0206020} {arXiv:hep-ph/0206020 [hep-ph]}
  \BibitemShut {NoStop}%
\bibitem [{\citenamefont {Schmaltz}\ and\ \citenamefont
  {Tucker-Smith}(2005)}]{Schmaltz:2005ky}%
  \BibitemOpen
  \bibfield  {author} {\bibinfo {author} {\bibfnamefont {M.}~\bibnamefont
  {Schmaltz}}\ and\ \bibinfo {author} {\bibfnamefont {D.}~\bibnamefont
  {Tucker-Smith}},\ }\href {\doibase 10.1146/annurev.nucl.55.090704.151502}
  {\bibfield  {journal} {\bibinfo  {journal} {Ann.Rev.Nucl.Part.Sci.}\ }\textbf
  {\bibinfo {volume} {55}},\ \bibinfo {pages} {229} (\bibinfo {year} {2005})},\
  \Eprint {http://arxiv.org/abs/hep-ph/0502182} {arXiv:hep-ph/0502182 [hep-ph]}
  \BibitemShut {NoStop}%
\bibitem [{\citenamefont {Arkani-Hamed}\ \emph {et~al.}(2001)\citenamefont
  {Arkani-Hamed}, \citenamefont {Cohen},\ and\ \citenamefont
  {Georgi}}]{ArkaniHamed:2001nc}%
  \BibitemOpen
  \bibfield  {author} {\bibinfo {author} {\bibfnamefont {N.}~\bibnamefont
  {Arkani-Hamed}}, \bibinfo {author} {\bibfnamefont {A.~G.}\ \bibnamefont
  {Cohen}}, \ and\ \bibinfo {author} {\bibfnamefont {H.}~\bibnamefont
  {Georgi}},\ }\href {\doibase 10.1016/S0370-2693(01)00741-9} {\bibfield
  {journal} {\bibinfo  {journal} {Phys.Lett.}\ }\textbf {\bibinfo {volume}
  {B513}},\ \bibinfo {pages} {232} (\bibinfo {year} {2001})},\ \Eprint
  {http://arxiv.org/abs/hep-ph/0105239} {arXiv:hep-ph/0105239 [hep-ph]}
  \BibitemShut {NoStop}%
\bibitem [{\citenamefont {Marzocca}\ \emph {et~al.}(2012)\citenamefont
  {Marzocca}, \citenamefont {Serone},\ and\ \citenamefont
  {Shu}}]{Marzocca:2012zn}%
  \BibitemOpen
  \bibfield  {author} {\bibinfo {author} {\bibfnamefont {D.}~\bibnamefont
  {Marzocca}}, \bibinfo {author} {\bibfnamefont {M.}~\bibnamefont {Serone}}, \
  and\ \bibinfo {author} {\bibfnamefont {J.}~\bibnamefont {Shu}},\ }\href
  {\doibase 10.1007/JHEP08(2012)013} {\bibfield  {journal} {\bibinfo  {journal}
  {JHEP}\ }\textbf {\bibinfo {volume} {08}},\ \bibinfo {pages} {013} (\bibinfo
  {year} {2012})},\ \Eprint {http://arxiv.org/abs/1205.0770} {arXiv:1205.0770
  [hep-ph]} \BibitemShut {NoStop}%
\bibitem [{\citenamefont {Ecker}\ \emph {et~al.}(1989)\citenamefont {Ecker},
  \citenamefont {Gasser}, \citenamefont {Leutwyler}, \citenamefont {Pich},\
  and\ \citenamefont {de~Rafael}}]{Ecker:1989yg}%
  \BibitemOpen
  \bibfield  {author} {\bibinfo {author} {\bibfnamefont {G.}~\bibnamefont
  {Ecker}}, \bibinfo {author} {\bibfnamefont {J.}~\bibnamefont {Gasser}},
  \bibinfo {author} {\bibfnamefont {H.}~\bibnamefont {Leutwyler}}, \bibinfo
  {author} {\bibfnamefont {A.}~\bibnamefont {Pich}}, \ and\ \bibinfo {author}
  {\bibfnamefont {E.}~\bibnamefont {de~Rafael}},\ }\href {\doibase
  10.1016/0370-2693(89)91627-4} {\bibfield  {journal} {\bibinfo  {journal}
  {Phys. Lett.}\ }\textbf {\bibinfo {volume} {B223}},\ \bibinfo {pages} {425}
  (\bibinfo {year} {1989})}\BibitemShut {NoStop}%
\bibitem [{\citenamefont {Giudice}\ \emph {et~al.}(2007)\citenamefont
  {Giudice}, \citenamefont {Grojean}, \citenamefont {Pomarol},\ and\
  \citenamefont {Rattazzi}}]{Giudice:2007fh}%
  \BibitemOpen
  \bibfield  {author} {\bibinfo {author} {\bibfnamefont {G.~F.}\ \bibnamefont
  {Giudice}}, \bibinfo {author} {\bibfnamefont {C.}~\bibnamefont {Grojean}},
  \bibinfo {author} {\bibfnamefont {A.}~\bibnamefont {Pomarol}}, \ and\
  \bibinfo {author} {\bibfnamefont {R.}~\bibnamefont {Rattazzi}},\ }\href
  {\doibase 10.1088/1126-6708/2007/06/045} {\bibfield  {journal} {\bibinfo
  {journal} {JHEP}\ }\textbf {\bibinfo {volume} {06}},\ \bibinfo {pages} {045}
  (\bibinfo {year} {2007})},\ \Eprint {http://arxiv.org/abs/hep-ph/0703164}
  {arXiv:hep-ph/0703164 [hep-ph]} \BibitemShut {NoStop}%
\bibitem [{\citenamefont {Barbieri}\ \emph {et~al.}(2015)\citenamefont
  {Barbieri}, \citenamefont {Greco}, \citenamefont {Rattazzi},\ and\
  \citenamefont {Wulzer}}]{Barbieri:2015lqa}%
  \BibitemOpen
  \bibfield  {author} {\bibinfo {author} {\bibfnamefont {R.}~\bibnamefont
  {Barbieri}}, \bibinfo {author} {\bibfnamefont {D.}~\bibnamefont {Greco}},
  \bibinfo {author} {\bibfnamefont {R.}~\bibnamefont {Rattazzi}}, \ and\
  \bibinfo {author} {\bibfnamefont {A.}~\bibnamefont {Wulzer}},\ }\href
  {\doibase 10.1007/JHEP08(2015)161} {\bibfield  {journal} {\bibinfo  {journal}
  {JHEP}\ }\textbf {\bibinfo {volume} {08}},\ \bibinfo {pages} {161} (\bibinfo
  {year} {2015})},\ \Eprint {http://arxiv.org/abs/1501.07803} {arXiv:1501.07803
  [hep-ph]} \BibitemShut {NoStop}%
\bibitem [{\citenamefont {Rattazzi}\ and\ \citenamefont
  {Zaffaroni}(2001)}]{Rattazzi:2000hs}%
  \BibitemOpen
  \bibfield  {author} {\bibinfo {author} {\bibfnamefont {R.}~\bibnamefont
  {Rattazzi}}\ and\ \bibinfo {author} {\bibfnamefont {A.}~\bibnamefont
  {Zaffaroni}},\ }\href {\doibase 10.1088/1126-6708/2001/04/021} {\bibfield
  {journal} {\bibinfo  {journal} {JHEP}\ }\textbf {\bibinfo {volume} {04}},\
  \bibinfo {pages} {021} (\bibinfo {year} {2001})},\ \Eprint
  {http://arxiv.org/abs/hep-th/0012248} {arXiv:hep-th/0012248 [hep-th]}
  \BibitemShut {NoStop}%
\bibitem [{\citenamefont {Low}(2015{\natexlab{a}})}]{Low:2014oga}%
  \BibitemOpen
  \bibfield  {author} {\bibinfo {author} {\bibfnamefont {I.}~\bibnamefont
  {Low}},\ }\href {\doibase 10.1103/PhysRevD.91.116005} {\bibfield  {journal}
  {\bibinfo  {journal} {Phys. Rev.}\ }\textbf {\bibinfo {volume} {D91}},\
  \bibinfo {pages} {116005} (\bibinfo {year} {2015}{\natexlab{a}})},\ \Eprint
  {http://arxiv.org/abs/1412.2146} {arXiv:1412.2146 [hep-ph]} \BibitemShut
  {NoStop}%
\bibitem [{\citenamefont {Low}(2015{\natexlab{b}})}]{Low:2015ogb}%
  \BibitemOpen
  \bibfield  {author} {\bibinfo {author} {\bibfnamefont {I.}~\bibnamefont
  {Low}},\ }\href@noop {} {\  (\bibinfo {year} {2015}{\natexlab{b}})},\ \Eprint
  {http://arxiv.org/abs/1512.01232} {arXiv:1512.01232 [hep-th]} \BibitemShut
  {NoStop}%
\bibitem [{\citenamefont {Cheng}\ and\ \citenamefont
  {Low}(2003)}]{Cheng:2003ju}%
  \BibitemOpen
  \bibfield  {author} {\bibinfo {author} {\bibfnamefont {H.-C.}\ \bibnamefont
  {Cheng}}\ and\ \bibinfo {author} {\bibfnamefont {I.}~\bibnamefont {Low}},\
  }\href {\doibase 10.1088/1126-6708/2003/09/051} {\bibfield  {journal}
  {\bibinfo  {journal} {JHEP}\ }\textbf {\bibinfo {volume} {09}},\ \bibinfo
  {pages} {051} (\bibinfo {year} {2003})},\ \Eprint
  {http://arxiv.org/abs/hep-ph/0308199} {arXiv:hep-ph/0308199 [hep-ph]}
  \BibitemShut {NoStop}%
\bibitem [{\citenamefont {Low}(2004)}]{Low:2004xc}%
  \BibitemOpen
  \bibfield  {author} {\bibinfo {author} {\bibfnamefont {I.}~\bibnamefont
  {Low}},\ }\href {\doibase 10.1088/1126-6708/2004/10/067} {\bibfield
  {journal} {\bibinfo  {journal} {JHEP}\ }\textbf {\bibinfo {volume} {10}},\
  \bibinfo {pages} {067} (\bibinfo {year} {2004})},\ \Eprint
  {http://arxiv.org/abs/hep-ph/0409025} {arXiv:hep-ph/0409025 [hep-ph]}
  \BibitemShut {NoStop}%
\bibitem [{\citenamefont {Contino}\ \emph
  {et~al.}(2007{\natexlab{a}})\citenamefont {Contino}, \citenamefont
  {Da~Rold},\ and\ \citenamefont {Pomarol}}]{Contino:2006qr}%
  \BibitemOpen
  \bibfield  {author} {\bibinfo {author} {\bibfnamefont {R.}~\bibnamefont
  {Contino}}, \bibinfo {author} {\bibfnamefont {L.}~\bibnamefont {Da~Rold}}, \
  and\ \bibinfo {author} {\bibfnamefont {A.}~\bibnamefont {Pomarol}},\ }\href
  {\doibase 10.1103/PhysRevD.75.055014} {\bibfield  {journal} {\bibinfo
  {journal} {Phys. Rev.}\ }\textbf {\bibinfo {volume} {D75}},\ \bibinfo {pages}
  {055014} (\bibinfo {year} {2007}{\natexlab{a}})},\ \Eprint
  {http://arxiv.org/abs/hep-ph/0612048} {arXiv:hep-ph/0612048 [hep-ph]}
  \BibitemShut {NoStop}%
\bibitem [{\citenamefont {Anastasiou}\ \emph {et~al.}(2009)\citenamefont
  {Anastasiou}, \citenamefont {Furlan},\ and\ \citenamefont
  {Santiago}}]{Anastasiou:2009rv}%
  \BibitemOpen
  \bibfield  {author} {\bibinfo {author} {\bibfnamefont {C.}~\bibnamefont
  {Anastasiou}}, \bibinfo {author} {\bibfnamefont {E.}~\bibnamefont {Furlan}},
  \ and\ \bibinfo {author} {\bibfnamefont {J.}~\bibnamefont {Santiago}},\
  }\href {\doibase 10.1103/PhysRevD.79.075003} {\bibfield  {journal} {\bibinfo
  {journal} {Phys. Rev.}\ }\textbf {\bibinfo {volume} {D79}},\ \bibinfo {pages}
  {075003} (\bibinfo {year} {2009})},\ \Eprint {http://arxiv.org/abs/0901.2117}
  {arXiv:0901.2117 [hep-ph]} \BibitemShut {NoStop}%
\bibitem [{\citenamefont {Medina}\ \emph {et~al.}(2007)\citenamefont {Medina},
  \citenamefont {Shah},\ and\ \citenamefont {Wagner}}]{Medina:2007hz}%
  \BibitemOpen
  \bibfield  {author} {\bibinfo {author} {\bibfnamefont {A.~D.}\ \bibnamefont
  {Medina}}, \bibinfo {author} {\bibfnamefont {N.~R.}\ \bibnamefont {Shah}}, \
  and\ \bibinfo {author} {\bibfnamefont {C.~E.~M.}\ \bibnamefont {Wagner}},\
  }\href {\doibase 10.1103/PhysRevD.76.095010} {\bibfield  {journal} {\bibinfo
  {journal} {Phys. Rev.}\ }\textbf {\bibinfo {volume} {D76}},\ \bibinfo {pages}
  {095010} (\bibinfo {year} {2007})},\ \Eprint {http://arxiv.org/abs/0706.1281}
  {arXiv:0706.1281 [hep-ph]} \BibitemShut {NoStop}%
\bibitem [{\citenamefont {Agashe}\ \emph {et~al.}(2005)\citenamefont {Agashe},
  \citenamefont {Contino},\ and\ \citenamefont {Pomarol}}]{Agashe:2004rs}%
  \BibitemOpen
  \bibfield  {author} {\bibinfo {author} {\bibfnamefont {K.}~\bibnamefont
  {Agashe}}, \bibinfo {author} {\bibfnamefont {R.}~\bibnamefont {Contino}}, \
  and\ \bibinfo {author} {\bibfnamefont {A.}~\bibnamefont {Pomarol}},\ }\href
  {\doibase 10.1016/j.nuclphysb.2005.04.035} {\bibfield  {journal} {\bibinfo
  {journal} {Nucl. Phys.}\ }\textbf {\bibinfo {volume} {B719}},\ \bibinfo
  {pages} {165} (\bibinfo {year} {2005})},\ \Eprint
  {http://arxiv.org/abs/hep-ph/0412089} {arXiv:hep-ph/0412089 [hep-ph]}
  \BibitemShut {NoStop}%
\bibitem [{\citenamefont {Panico}\ and\ \citenamefont
  {Wulzer}(2011)}]{Panico:2011pw}%
  \BibitemOpen
  \bibfield  {author} {\bibinfo {author} {\bibfnamefont {G.}~\bibnamefont
  {Panico}}\ and\ \bibinfo {author} {\bibfnamefont {A.}~\bibnamefont
  {Wulzer}},\ }\href {\doibase 10.1007/JHEP09(2011)135} {\bibfield  {journal}
  {\bibinfo  {journal} {JHEP}\ }\textbf {\bibinfo {volume} {09}},\ \bibinfo
  {pages} {135} (\bibinfo {year} {2011})},\ \Eprint
  {http://arxiv.org/abs/1106.2719} {arXiv:1106.2719 [hep-ph]} \BibitemShut
  {NoStop}%
\bibitem [{\citenamefont {De~Curtis}\ \emph {et~al.}(2012)\citenamefont
  {De~Curtis}, \citenamefont {Redi},\ and\ \citenamefont
  {Tesi}}]{DeCurtis:2011yx}%
  \BibitemOpen
  \bibfield  {author} {\bibinfo {author} {\bibfnamefont {S.}~\bibnamefont
  {De~Curtis}}, \bibinfo {author} {\bibfnamefont {M.}~\bibnamefont {Redi}}, \
  and\ \bibinfo {author} {\bibfnamefont {A.}~\bibnamefont {Tesi}},\ }\href
  {\doibase 10.1007/JHEP04(2012)042} {\bibfield  {journal} {\bibinfo  {journal}
  {JHEP}\ }\textbf {\bibinfo {volume} {04}},\ \bibinfo {pages} {042} (\bibinfo
  {year} {2012})},\ \Eprint {http://arxiv.org/abs/1110.1613} {arXiv:1110.1613
  [hep-ph]} \BibitemShut {NoStop}%
\bibitem [{\citenamefont {Grossman}\ and\ \citenamefont
  {Neubert}(2000)}]{Grossman:1999ra}%
  \BibitemOpen
  \bibfield  {author} {\bibinfo {author} {\bibfnamefont {Y.}~\bibnamefont
  {Grossman}}\ and\ \bibinfo {author} {\bibfnamefont {M.}~\bibnamefont
  {Neubert}},\ }\href {\doibase 10.1016/S0370-2693(00)00054-X} {\bibfield
  {journal} {\bibinfo  {journal} {Phys. Lett.}\ }\textbf {\bibinfo {volume}
  {B474}},\ \bibinfo {pages} {361} (\bibinfo {year} {2000})},\ \Eprint
  {http://arxiv.org/abs/hep-ph/9912408} {arXiv:hep-ph/9912408 [hep-ph]}
  \BibitemShut {NoStop}%
\bibitem [{\citenamefont {Gripaios}\ \emph {et~al.}(2009)\citenamefont
  {Gripaios}, \citenamefont {Pomarol}, \citenamefont {Riva},\ and\
  \citenamefont {Serra}}]{Gripaios:2009pe}%
  \BibitemOpen
  \bibfield  {author} {\bibinfo {author} {\bibfnamefont {B.}~\bibnamefont
  {Gripaios}}, \bibinfo {author} {\bibfnamefont {A.}~\bibnamefont {Pomarol}},
  \bibinfo {author} {\bibfnamefont {F.}~\bibnamefont {Riva}}, \ and\ \bibinfo
  {author} {\bibfnamefont {J.}~\bibnamefont {Serra}},\ }\href {\doibase
  10.1088/1126-6708/2009/04/070} {\bibfield  {journal} {\bibinfo  {journal}
  {JHEP}\ }\textbf {\bibinfo {volume} {04}},\ \bibinfo {pages} {070} (\bibinfo
  {year} {2009})},\ \Eprint {http://arxiv.org/abs/0902.1483} {arXiv:0902.1483
  [hep-ph]} \BibitemShut {NoStop}%
\bibitem [{\citenamefont {Mrazek}\ \emph {et~al.}(2011)\citenamefont {Mrazek},
  \citenamefont {Pomarol}, \citenamefont {Rattazzi}, \citenamefont {Redi},
  \citenamefont {Serra},\ and\ \citenamefont {Wulzer}}]{Mrazek:2011iu}%
  \BibitemOpen
  \bibfield  {author} {\bibinfo {author} {\bibfnamefont {J.}~\bibnamefont
  {Mrazek}}, \bibinfo {author} {\bibfnamefont {A.}~\bibnamefont {Pomarol}},
  \bibinfo {author} {\bibfnamefont {R.}~\bibnamefont {Rattazzi}}, \bibinfo
  {author} {\bibfnamefont {M.}~\bibnamefont {Redi}}, \bibinfo {author}
  {\bibfnamefont {J.}~\bibnamefont {Serra}}, \ and\ \bibinfo {author}
  {\bibfnamefont {A.}~\bibnamefont {Wulzer}},\ }\href {\doibase
  10.1016/j.nuclphysb.2011.07.008} {\bibfield  {journal} {\bibinfo  {journal}
  {Nucl. Phys.}\ }\textbf {\bibinfo {volume} {B853}},\ \bibinfo {pages} {1}
  (\bibinfo {year} {2011})},\ \Eprint {http://arxiv.org/abs/1105.5403}
  {arXiv:1105.5403 [hep-ph]} \BibitemShut {NoStop}%
\bibitem [{\citenamefont {Hall}\ \emph {et~al.}(2002)\citenamefont {Hall},
  \citenamefont {Nomura},\ and\ \citenamefont {Tucker-Smith}}]{Hall:2001zb}%
  \BibitemOpen
  \bibfield  {author} {\bibinfo {author} {\bibfnamefont {L.~J.}\ \bibnamefont
  {Hall}}, \bibinfo {author} {\bibfnamefont {Y.}~\bibnamefont {Nomura}}, \ and\
  \bibinfo {author} {\bibfnamefont {D.}~\bibnamefont {Tucker-Smith}},\ }\href
  {\doibase 10.1016/S0550-3213(02)00539-4} {\bibfield  {journal} {\bibinfo
  {journal} {Nucl. Phys.}\ }\textbf {\bibinfo {volume} {B639}},\ \bibinfo
  {pages} {307} (\bibinfo {year} {2002})},\ \Eprint
  {http://arxiv.org/abs/hep-ph/0107331} {arXiv:hep-ph/0107331 [hep-ph]}
  \BibitemShut {NoStop}%
\bibitem [{\citenamefont {Kubo}\ \emph {et~al.}(2002)\citenamefont {Kubo},
  \citenamefont {Lim},\ and\ \citenamefont {Yamashita}}]{Kubo:2001zc}%
  \BibitemOpen
  \bibfield  {author} {\bibinfo {author} {\bibfnamefont {M.}~\bibnamefont
  {Kubo}}, \bibinfo {author} {\bibfnamefont {C.~S.}\ \bibnamefont {Lim}}, \
  and\ \bibinfo {author} {\bibfnamefont {H.}~\bibnamefont {Yamashita}},\ }\href
  {\doibase 10.1142/S0217732302008988} {\bibfield  {journal} {\bibinfo
  {journal} {Mod. Phys. Lett.}\ }\textbf {\bibinfo {volume} {A17}},\ \bibinfo
  {pages} {2249} (\bibinfo {year} {2002})},\ \Eprint
  {http://arxiv.org/abs/hep-ph/0111327} {arXiv:hep-ph/0111327 [hep-ph]}
  \BibitemShut {NoStop}%
\bibitem [{\citenamefont {Burdman}\ and\ \citenamefont
  {Nomura}(2003)}]{Burdman:2002se}%
  \BibitemOpen
  \bibfield  {author} {\bibinfo {author} {\bibfnamefont {G.}~\bibnamefont
  {Burdman}}\ and\ \bibinfo {author} {\bibfnamefont {Y.}~\bibnamefont
  {Nomura}},\ }\href {\doibase 10.1016/S0550-3213(03)00088-9} {\bibfield
  {journal} {\bibinfo  {journal} {Nucl. Phys.}\ }\textbf {\bibinfo {volume}
  {B656}},\ \bibinfo {pages} {3} (\bibinfo {year} {2003})},\ \Eprint
  {http://arxiv.org/abs/hep-ph/0210257} {arXiv:hep-ph/0210257 [hep-ph]}
  \BibitemShut {NoStop}%
\bibitem [{\citenamefont {Scrucca}\ \emph {et~al.}(2003)\citenamefont
  {Scrucca}, \citenamefont {Serone},\ and\ \citenamefont
  {Silvestrini}}]{Scrucca:2003ra}%
  \BibitemOpen
  \bibfield  {author} {\bibinfo {author} {\bibfnamefont {C.~A.}\ \bibnamefont
  {Scrucca}}, \bibinfo {author} {\bibfnamefont {M.}~\bibnamefont {Serone}}, \
  and\ \bibinfo {author} {\bibfnamefont {L.}~\bibnamefont {Silvestrini}},\
  }\href {\doibase 10.1016/j.nuclphysb.2003.07.013} {\bibfield  {journal}
  {\bibinfo  {journal} {Nucl. Phys.}\ }\textbf {\bibinfo {volume} {B669}},\
  \bibinfo {pages} {128} (\bibinfo {year} {2003})},\ \Eprint
  {http://arxiv.org/abs/hep-ph/0304220} {arXiv:hep-ph/0304220 [hep-ph]}
  \BibitemShut {NoStop}%
\bibitem [{\citenamefont {Contino}\ \emph {et~al.}(2003)\citenamefont
  {Contino}, \citenamefont {Nomura},\ and\ \citenamefont
  {Pomarol}}]{Contino:2003ve}%
  \BibitemOpen
  \bibfield  {author} {\bibinfo {author} {\bibfnamefont {R.}~\bibnamefont
  {Contino}}, \bibinfo {author} {\bibfnamefont {Y.}~\bibnamefont {Nomura}}, \
  and\ \bibinfo {author} {\bibfnamefont {A.}~\bibnamefont {Pomarol}},\ }\href
  {\doibase 10.1016/j.nuclphysb.2003.08.027} {\bibfield  {journal} {\bibinfo
  {journal} {Nucl. Phys.}\ }\textbf {\bibinfo {volume} {B671}},\ \bibinfo
  {pages} {148} (\bibinfo {year} {2003})},\ \Eprint
  {http://arxiv.org/abs/hep-ph/0306259} {arXiv:hep-ph/0306259 [hep-ph]}
  \BibitemShut {NoStop}%
\bibitem [{\citenamefont {Panico}\ \emph {et~al.}(2006)\citenamefont {Panico},
  \citenamefont {Serone},\ and\ \citenamefont {Wulzer}}]{Panico:2005dh}%
  \BibitemOpen
  \bibfield  {author} {\bibinfo {author} {\bibfnamefont {G.}~\bibnamefont
  {Panico}}, \bibinfo {author} {\bibfnamefont {M.}~\bibnamefont {Serone}}, \
  and\ \bibinfo {author} {\bibfnamefont {A.}~\bibnamefont {Wulzer}},\ }\href
  {\doibase 10.1016/j.nuclphysb.2006.01.025} {\bibfield  {journal} {\bibinfo
  {journal} {Nucl. Phys.}\ }\textbf {\bibinfo {volume} {B739}},\ \bibinfo
  {pages} {186} (\bibinfo {year} {2006})},\ \Eprint
  {http://arxiv.org/abs/hep-ph/0510373} {arXiv:hep-ph/0510373 [hep-ph]}
  \BibitemShut {NoStop}%
\bibitem [{\citenamefont {Panico}\ \emph {et~al.}(2007)\citenamefont {Panico},
  \citenamefont {Serone},\ and\ \citenamefont {Wulzer}}]{Panico:2006em}%
  \BibitemOpen
  \bibfield  {author} {\bibinfo {author} {\bibfnamefont {G.}~\bibnamefont
  {Panico}}, \bibinfo {author} {\bibfnamefont {M.}~\bibnamefont {Serone}}, \
  and\ \bibinfo {author} {\bibfnamefont {A.}~\bibnamefont {Wulzer}},\ }\href
  {\doibase 10.1016/j.nuclphysb.2006.10.032} {\bibfield  {journal} {\bibinfo
  {journal} {Nucl. Phys.}\ }\textbf {\bibinfo {volume} {B762}},\ \bibinfo
  {pages} {189} (\bibinfo {year} {2007})},\ \Eprint
  {http://arxiv.org/abs/hep-ph/0605292} {arXiv:hep-ph/0605292 [hep-ph]}
  \BibitemShut {NoStop}%
\bibitem [{\citenamefont {Vignaroli}(2012)}]{Vignaroli:2012sf}%
  \BibitemOpen
  \bibfield  {author} {\bibinfo {author} {\bibfnamefont {N.}~\bibnamefont
  {Vignaroli}},\ }\href {\doibase 10.1007/JHEP07(2012)158} {\bibfield
  {journal} {\bibinfo  {journal} {JHEP}\ }\textbf {\bibinfo {volume} {07}},\
  \bibinfo {pages} {158} (\bibinfo {year} {2012})},\ \Eprint
  {http://arxiv.org/abs/1204.0468} {arXiv:1204.0468 [hep-ph]} \BibitemShut
  {NoStop}%
\bibitem [{\citenamefont {De~Simone}\ \emph {et~al.}(2013)\citenamefont
  {De~Simone}, \citenamefont {Matsedonskyi}, \citenamefont {Rattazzi},\ and\
  \citenamefont {Wulzer}}]{DeSimone:2012fs}%
  \BibitemOpen
  \bibfield  {author} {\bibinfo {author} {\bibfnamefont {A.}~\bibnamefont
  {De~Simone}}, \bibinfo {author} {\bibfnamefont {O.}~\bibnamefont
  {Matsedonskyi}}, \bibinfo {author} {\bibfnamefont {R.}~\bibnamefont
  {Rattazzi}}, \ and\ \bibinfo {author} {\bibfnamefont {A.}~\bibnamefont
  {Wulzer}},\ }\href {\doibase 10.1007/JHEP04(2013)004} {\bibfield  {journal}
  {\bibinfo  {journal} {JHEP}\ }\textbf {\bibinfo {volume} {04}},\ \bibinfo
  {pages} {004} (\bibinfo {year} {2013})},\ \Eprint
  {http://arxiv.org/abs/1211.5663} {arXiv:1211.5663 [hep-ph]} \BibitemShut
  {NoStop}%
\bibitem [{\citenamefont {Delaunay}\ \emph {et~al.}(2013)\citenamefont
  {Delaunay}, \citenamefont {Grojean},\ and\ \citenamefont
  {Perez}}]{Delaunay:2013iia}%
  \BibitemOpen
  \bibfield  {author} {\bibinfo {author} {\bibfnamefont {C.}~\bibnamefont
  {Delaunay}}, \bibinfo {author} {\bibfnamefont {C.}~\bibnamefont {Grojean}}, \
  and\ \bibinfo {author} {\bibfnamefont {G.}~\bibnamefont {Perez}},\ }\href
  {\doibase 10.1007/JHEP09(2013)090} {\bibfield  {journal} {\bibinfo  {journal}
  {JHEP}\ }\textbf {\bibinfo {volume} {09}},\ \bibinfo {pages} {090} (\bibinfo
  {year} {2013})},\ \Eprint {http://arxiv.org/abs/1303.5701} {arXiv:1303.5701
  [hep-ph]} \BibitemShut {NoStop}%
\bibitem [{\citenamefont {Gillioz}\ \emph {et~al.}(2014)\citenamefont
  {Gillioz}, \citenamefont {Gröber}, \citenamefont {Kapuvari},\ and\
  \citenamefont {Mühlleitner}}]{Gillioz:2013pba}%
  \BibitemOpen
  \bibfield  {author} {\bibinfo {author} {\bibfnamefont {M.}~\bibnamefont
  {Gillioz}}, \bibinfo {author} {\bibfnamefont {R.}~\bibnamefont {Gröber}},
  \bibinfo {author} {\bibfnamefont {A.}~\bibnamefont {Kapuvari}}, \ and\
  \bibinfo {author} {\bibfnamefont {M.}~\bibnamefont {Mühlleitner}},\ }\href
  {\doibase 10.1007/JHEP03(2014)037} {\bibfield  {journal} {\bibinfo  {journal}
  {JHEP}\ }\textbf {\bibinfo {volume} {03}},\ \bibinfo {pages} {037} (\bibinfo
  {year} {2014})},\ \Eprint {http://arxiv.org/abs/1311.4453} {arXiv:1311.4453
  [hep-ph]} \BibitemShut {NoStop}%
\bibitem [{\citenamefont {Han}\ \emph {et~al.}(2003)\citenamefont {Han},
  \citenamefont {Logan}, \citenamefont {McElrath},\ and\ \citenamefont
  {Wang}}]{Han:2003wu}%
  \BibitemOpen
  \bibfield  {author} {\bibinfo {author} {\bibfnamefont {T.}~\bibnamefont
  {Han}}, \bibinfo {author} {\bibfnamefont {H.~E.}\ \bibnamefont {Logan}},
  \bibinfo {author} {\bibfnamefont {B.}~\bibnamefont {McElrath}}, \ and\
  \bibinfo {author} {\bibfnamefont {L.-T.}\ \bibnamefont {Wang}},\ }\href
  {\doibase 10.1103/PhysRevD.67.095004} {\bibfield  {journal} {\bibinfo
  {journal} {Phys. Rev.}\ }\textbf {\bibinfo {volume} {D67}},\ \bibinfo {pages}
  {095004} (\bibinfo {year} {2003})},\ \Eprint
  {http://arxiv.org/abs/hep-ph/0301040} {arXiv:hep-ph/0301040 [hep-ph]}
  \BibitemShut {NoStop}%
\bibitem [{\citenamefont {Carena}\ \emph {et~al.}(2007)\citenamefont {Carena},
  \citenamefont {Hubisz}, \citenamefont {Perelstein},\ and\ \citenamefont
  {Verdier}}]{Carena:2006jx}%
  \BibitemOpen
  \bibfield  {author} {\bibinfo {author} {\bibfnamefont {M.}~\bibnamefont
  {Carena}}, \bibinfo {author} {\bibfnamefont {J.}~\bibnamefont {Hubisz}},
  \bibinfo {author} {\bibfnamefont {M.}~\bibnamefont {Perelstein}}, \ and\
  \bibinfo {author} {\bibfnamefont {P.}~\bibnamefont {Verdier}},\ }\href
  {\doibase 10.1103/PhysRevD.75.091701} {\bibfield  {journal} {\bibinfo
  {journal} {Phys. Rev.}\ }\textbf {\bibinfo {volume} {D75}},\ \bibinfo {pages}
  {091701} (\bibinfo {year} {2007})},\ \Eprint
  {http://arxiv.org/abs/hep-ph/0610156} {arXiv:hep-ph/0610156 [hep-ph]}
  \BibitemShut {NoStop}%
\bibitem [{\citenamefont {Matsumoto}\ \emph {et~al.}(2008)\citenamefont
  {Matsumoto}, \citenamefont {Moroi},\ and\ \citenamefont
  {Tobe}}]{Matsumoto:2008fq}%
  \BibitemOpen
  \bibfield  {author} {\bibinfo {author} {\bibfnamefont {S.}~\bibnamefont
  {Matsumoto}}, \bibinfo {author} {\bibfnamefont {T.}~\bibnamefont {Moroi}}, \
  and\ \bibinfo {author} {\bibfnamefont {K.}~\bibnamefont {Tobe}},\ }\href
  {\doibase 10.1103/PhysRevD.78.055018} {\bibfield  {journal} {\bibinfo
  {journal} {Phys. Rev.}\ }\textbf {\bibinfo {volume} {D78}},\ \bibinfo {pages}
  {055018} (\bibinfo {year} {2008})},\ \Eprint {http://arxiv.org/abs/0806.3837}
  {arXiv:0806.3837 [hep-ph]} \BibitemShut {NoStop}%
\bibitem [{\citenamefont {Contino}\ and\ \citenamefont
  {Salvarezza}(2015)}]{Contino:2015gdp}%
  \BibitemOpen
  \bibfield  {author} {\bibinfo {author} {\bibfnamefont {R.}~\bibnamefont
  {Contino}}\ and\ \bibinfo {author} {\bibfnamefont {M.}~\bibnamefont
  {Salvarezza}},\ }\href {\doibase 10.1103/PhysRevD.92.115010} {\bibfield
  {journal} {\bibinfo  {journal} {Phys. Rev.}\ }\textbf {\bibinfo {volume}
  {D92}},\ \bibinfo {pages} {115010} (\bibinfo {year} {2015})},\ \Eprint
  {http://arxiv.org/abs/1511.00592} {arXiv:1511.00592 [hep-ph]} \BibitemShut
  {NoStop}%
\bibitem [{\citenamefont {Huber}\ and\ \citenamefont
  {Shafi}(2001)}]{Huber:2000ie}%
  \BibitemOpen
  \bibfield  {author} {\bibinfo {author} {\bibfnamefont {S.~J.}\ \bibnamefont
  {Huber}}\ and\ \bibinfo {author} {\bibfnamefont {Q.}~\bibnamefont {Shafi}},\
  }\href {\doibase 10.1016/S0370-2693(00)01399-X} {\bibfield  {journal}
  {\bibinfo  {journal} {Phys. Lett.}\ }\textbf {\bibinfo {volume} {B498}},\
  \bibinfo {pages} {256} (\bibinfo {year} {2001})},\ \Eprint
  {http://arxiv.org/abs/hep-ph/0010195} {arXiv:hep-ph/0010195 [hep-ph]}
  \BibitemShut {NoStop}%
\bibitem [{\citenamefont {Matsedonskyi}\ \emph {et~al.}(2013)\citenamefont
  {Matsedonskyi}, \citenamefont {Panico},\ and\ \citenamefont
  {Wulzer}}]{Matsedonskyi:2012ym}%
  \BibitemOpen
  \bibfield  {author} {\bibinfo {author} {\bibfnamefont {O.}~\bibnamefont
  {Matsedonskyi}}, \bibinfo {author} {\bibfnamefont {G.}~\bibnamefont
  {Panico}}, \ and\ \bibinfo {author} {\bibfnamefont {A.}~\bibnamefont
  {Wulzer}},\ }\href {\doibase 10.1007/JHEP01(2013)164} {\bibfield  {journal}
  {\bibinfo  {journal} {JHEP}\ }\textbf {\bibinfo {volume} {01}},\ \bibinfo
  {pages} {164} (\bibinfo {year} {2013})},\ \Eprint
  {http://arxiv.org/abs/1204.6333} {arXiv:1204.6333 [hep-ph]} \BibitemShut
  {NoStop}%
\bibitem [{\citenamefont {McGuirk}\ \emph {et~al.}(2010)\citenamefont
  {McGuirk}, \citenamefont {Shiu},\ and\ \citenamefont
  {Sumitomo}}]{McGuirk:2009am}%
  \BibitemOpen
  \bibfield  {author} {\bibinfo {author} {\bibfnamefont {P.}~\bibnamefont
  {McGuirk}}, \bibinfo {author} {\bibfnamefont {G.}~\bibnamefont {Shiu}}, \
  and\ \bibinfo {author} {\bibfnamefont {Y.}~\bibnamefont {Sumitomo}},\ }\href
  {\doibase 10.1103/PhysRevD.81.026005} {\bibfield  {journal} {\bibinfo
  {journal} {Phys. Rev.}\ }\textbf {\bibinfo {volume} {D81}},\ \bibinfo {pages}
  {026005} (\bibinfo {year} {2010})},\ \Eprint {http://arxiv.org/abs/0911.0019}
  {arXiv:0911.0019 [hep-th]} \BibitemShut {NoStop}%
\bibitem [{\citenamefont {Gherghetta}\ and\ \citenamefont
  {Pomarol}(2000)}]{Gherghetta:2000qt}%
  \BibitemOpen
  \bibfield  {author} {\bibinfo {author} {\bibfnamefont {T.}~\bibnamefont
  {Gherghetta}}\ and\ \bibinfo {author} {\bibfnamefont {A.}~\bibnamefont
  {Pomarol}},\ }\href {\doibase 10.1016/S0550-3213(00)00392-8} {\bibfield
  {journal} {\bibinfo  {journal} {Nucl. Phys.}\ }\textbf {\bibinfo {volume}
  {B586}},\ \bibinfo {pages} {141} (\bibinfo {year} {2000})},\ \Eprint
  {http://arxiv.org/abs/hep-ph/0003129} {arXiv:hep-ph/0003129 [hep-ph]}
  \BibitemShut {NoStop}%
\bibitem [{\citenamefont {Contino}\ \emph
  {et~al.}(2007{\natexlab{b}})\citenamefont {Contino}, \citenamefont {Kramer},
  \citenamefont {Son},\ and\ \citenamefont {Sundrum}}]{Contino:2006nn}%
  \BibitemOpen
  \bibfield  {author} {\bibinfo {author} {\bibfnamefont {R.}~\bibnamefont
  {Contino}}, \bibinfo {author} {\bibfnamefont {T.}~\bibnamefont {Kramer}},
  \bibinfo {author} {\bibfnamefont {M.}~\bibnamefont {Son}}, \ and\ \bibinfo
  {author} {\bibfnamefont {R.}~\bibnamefont {Sundrum}},\ }\href {\doibase
  10.1088/1126-6708/2007/05/074} {\bibfield  {journal} {\bibinfo  {journal}
  {JHEP}\ }\textbf {\bibinfo {volume} {05}},\ \bibinfo {pages} {074} (\bibinfo
  {year} {2007}{\natexlab{b}})},\ \Eprint {http://arxiv.org/abs/hep-ph/0612180}
  {arXiv:hep-ph/0612180 [hep-ph]} \BibitemShut {NoStop}%
\bibitem [{\citenamefont {Serone}(2010)}]{Serone:2009kf}%
  \BibitemOpen
  \bibfield  {author} {\bibinfo {author} {\bibfnamefont {M.}~\bibnamefont
  {Serone}},\ }\href {\doibase 10.1088/1367-2630/12/7/075013} {\bibfield
  {journal} {\bibinfo  {journal} {New J. Phys.}\ }\textbf {\bibinfo {volume}
  {12}},\ \bibinfo {pages} {075013} (\bibinfo {year} {2010})},\ \Eprint
  {http://arxiv.org/abs/0909.5619} {arXiv:0909.5619 [hep-ph]} \BibitemShut
  {NoStop}%
\bibitem [{\citenamefont {Seiberg}(2006)}]{Seiberg:2006wf}%
  \BibitemOpen
  \bibfield  {author} {\bibinfo {author} {\bibfnamefont {N.}~\bibnamefont
  {Seiberg}},\ }in\ \href@noop {} {\emph {\bibinfo {booktitle} {{The Quantum
  Structure of Space and Time}}}}\ (\bibinfo {year} {2006})\ pp.\ \bibinfo
  {pages} {163--178},\ \Eprint {http://arxiv.org/abs/hep-th/0601234}
  {arXiv:hep-th/0601234 [hep-th]} \BibitemShut {NoStop}%
\bibitem [{\citenamefont {You}\ and\ \citenamefont {Xu}(2015)}]{You:2014vea}%
  \BibitemOpen
  \bibfield  {author} {\bibinfo {author} {\bibfnamefont {Y.-Z.}\ \bibnamefont
  {You}}\ and\ \bibinfo {author} {\bibfnamefont {C.}~\bibnamefont {Xu}},\
  }\href {\doibase 10.1103/PhysRevB.91.125147} {\bibfield  {journal} {\bibinfo
  {journal} {Phys. Rev.}\ }\textbf {\bibinfo {volume} {B91}},\ \bibinfo {pages}
  {125147} (\bibinfo {year} {2015})},\ \Eprint {http://arxiv.org/abs/1412.4784}
  {arXiv:1412.4784 [cond-mat.str-el]} \BibitemShut {NoStop}%
\bibitem [{\citenamefont {Kaplan}\ and\ \citenamefont
  {Sun}(2012)}]{Kaplan:2011vz}%
  \BibitemOpen
  \bibfield  {author} {\bibinfo {author} {\bibfnamefont {D.~B.}\ \bibnamefont
  {Kaplan}}\ and\ \bibinfo {author} {\bibfnamefont {S.}~\bibnamefont {Sun}},\
  }\href {\doibase 10.1103/PhysRevLett.108.181807,
  10.1103/PhysRevLett.108.209901} {\bibfield  {journal} {\bibinfo  {journal}
  {Phys.Rev.Lett.}\ }\textbf {\bibinfo {volume} {108}},\ \bibinfo {pages}
  {181807} (\bibinfo {year} {2012})},\ \Eprint {http://arxiv.org/abs/1112.0302}
  {arXiv:1112.0302 [hep-ph]} \BibitemShut {NoStop}%
\bibitem [{\citenamefont {Grabowska}\ and\ \citenamefont
  {Kaplan}(2015)}]{Grabowska:2015rda}%
  \BibitemOpen
  \bibfield  {author} {\bibinfo {author} {\bibfnamefont {D.}~\bibnamefont
  {Grabowska}}\ and\ \bibinfo {author} {\bibfnamefont {D.~B.}\ \bibnamefont
  {Kaplan}},\ }\href@noop {} {\  (\bibinfo {year} {2015})},\ \Eprint
  {http://arxiv.org/abs/1509.05758} {arXiv:1509.05758 [hep-ph]} \BibitemShut
  {NoStop}%
\bibitem [{\citenamefont {Descotes-Genon}\ \emph {et~al.}(2013)\citenamefont
  {Descotes-Genon}, \citenamefont {Matias},\ and\ \citenamefont
  {Virto}}]{Descotes-Genon:2013wba}%
  \BibitemOpen
  \bibfield  {author} {\bibinfo {author} {\bibfnamefont {S.}~\bibnamefont
  {Descotes-Genon}}, \bibinfo {author} {\bibfnamefont {J.}~\bibnamefont
  {Matias}}, \ and\ \bibinfo {author} {\bibfnamefont {J.}~\bibnamefont
  {Virto}},\ }\href {\doibase 10.1103/PhysRevD.88.074002} {\bibfield  {journal}
  {\bibinfo  {journal} {Phys.Rev.}\ }\textbf {\bibinfo {volume} {D88}},\
  \bibinfo {pages} {074002} (\bibinfo {year} {2013})},\ \Eprint
  {http://arxiv.org/abs/1307.5683} {arXiv:1307.5683 [hep-ph]} \BibitemShut
  {NoStop}%
\bibitem [{\citenamefont {Buras}\ and\ \citenamefont
  {Girrbach}(2013)}]{Buras:2013qja}%
  \BibitemOpen
  \bibfield  {author} {\bibinfo {author} {\bibfnamefont {A.~J.}\ \bibnamefont
  {Buras}}\ and\ \bibinfo {author} {\bibfnamefont {J.}~\bibnamefont
  {Girrbach}},\ }\href {\doibase 10.1007/JHEP12(2013)009} {\bibfield  {journal}
  {\bibinfo  {journal} {JHEP}\ }\textbf {\bibinfo {volume} {1312}},\ \bibinfo
  {pages} {009} (\bibinfo {year} {2013})},\ \Eprint
  {http://arxiv.org/abs/1309.2466} {arXiv:1309.2466 [hep-ph]} \BibitemShut
  {NoStop}%
\bibitem [{\citenamefont {Altmannshofer}\ and\ \citenamefont
  {Straub}(2013)}]{Altmannshofer:2013foa}%
  \BibitemOpen
  \bibfield  {author} {\bibinfo {author} {\bibfnamefont {W.}~\bibnamefont
  {Altmannshofer}}\ and\ \bibinfo {author} {\bibfnamefont {D.~M.}\ \bibnamefont
  {Straub}},\ }\href {\doibase 10.1140/epjc/s10052-013-2646-9} {\bibfield
  {journal} {\bibinfo  {journal} {Eur.Phys.J.}\ }\textbf {\bibinfo {volume}
  {C73}},\ \bibinfo {pages} {2646} (\bibinfo {year} {2013})},\ \Eprint
  {http://arxiv.org/abs/1308.1501} {arXiv:1308.1501 [hep-ph]} \BibitemShut
  {NoStop}%
\bibitem [{\citenamefont {Gauld}\ \emph {et~al.}(2014)\citenamefont {Gauld},
  \citenamefont {Goertz},\ and\ \citenamefont {Haisch}}]{Gauld:2013qja}%
  \BibitemOpen
  \bibfield  {author} {\bibinfo {author} {\bibfnamefont {R.}~\bibnamefont
  {Gauld}}, \bibinfo {author} {\bibfnamefont {F.}~\bibnamefont {Goertz}}, \
  and\ \bibinfo {author} {\bibfnamefont {U.}~\bibnamefont {Haisch}},\ }\href
  {\doibase 10.1007/JHEP01(2014)069} {\bibfield  {journal} {\bibinfo  {journal}
  {JHEP}\ }\textbf {\bibinfo {volume} {1401}},\ \bibinfo {pages} {069}
  (\bibinfo {year} {2014})},\ \Eprint {http://arxiv.org/abs/1310.1082}
  {arXiv:1310.1082 [hep-ph]} \BibitemShut {NoStop}%
\bibitem [{\citenamefont {Aad}\ \emph {et~al.}(2015{\natexlab{a}})\citenamefont
  {Aad} \emph {et~al.}}]{Aad:2015owa}%
  \BibitemOpen
  \bibfield  {author} {\bibinfo {author} {\bibfnamefont {G.}~\bibnamefont
  {Aad}} \emph {et~al.} (\bibinfo {collaboration} {ATLAS}),\ }\href {\doibase
  10.1007/JHEP12(2015)055} {\bibfield  {journal} {\bibinfo  {journal} {JHEP}\
  }\textbf {\bibinfo {volume} {12}},\ \bibinfo {pages} {055} (\bibinfo {year}
  {2015}{\natexlab{a}})},\ \Eprint {http://arxiv.org/abs/1506.00962}
  {arXiv:1506.00962 [hep-ex]} \BibitemShut {NoStop}%
\bibitem [{\citenamefont {Hisano}\ \emph {et~al.}(2015)\citenamefont {Hisano},
  \citenamefont {Nagata},\ and\ \citenamefont {Omura}}]{Hisano:2015gna}%
  \BibitemOpen
  \bibfield  {author} {\bibinfo {author} {\bibfnamefont {J.}~\bibnamefont
  {Hisano}}, \bibinfo {author} {\bibfnamefont {N.}~\bibnamefont {Nagata}}, \
  and\ \bibinfo {author} {\bibfnamefont {Y.}~\bibnamefont {Omura}},\ }\href
  {\doibase 10.1103/PhysRevD.92.055001} {\bibfield  {journal} {\bibinfo
  {journal} {Phys. Rev.}\ }\textbf {\bibinfo {volume} {D92}},\ \bibinfo {pages}
  {055001} (\bibinfo {year} {2015})},\ \Eprint
  {http://arxiv.org/abs/1506.03931} {arXiv:1506.03931 [hep-ph]} \BibitemShut
  {NoStop}%
\bibitem [{\citenamefont {Dobrescu}\ and\ \citenamefont
  {Liu}(2015)}]{Dobrescu:2015qna}%
  \BibitemOpen
  \bibfield  {author} {\bibinfo {author} {\bibfnamefont {B.~A.}\ \bibnamefont
  {Dobrescu}}\ and\ \bibinfo {author} {\bibfnamefont {Z.}~\bibnamefont {Liu}},\
  }\href {\doibase 10.1103/PhysRevLett.115.211802} {\bibfield  {journal}
  {\bibinfo  {journal} {Phys. Rev. Lett.}\ }\textbf {\bibinfo {volume} {115}},\
  \bibinfo {pages} {211802} (\bibinfo {year} {2015})},\ \Eprint
  {http://arxiv.org/abs/1506.06736} {arXiv:1506.06736 [hep-ph]} \BibitemShut
  {NoStop}%
\bibitem [{\citenamefont {Brehmer}\ \emph {et~al.}(2015)\citenamefont
  {Brehmer}, \citenamefont {Hewett}, \citenamefont {Kopp}, \citenamefont
  {Rizzo},\ and\ \citenamefont {Tattersall}}]{Brehmer:2015cia}%
  \BibitemOpen
  \bibfield  {author} {\bibinfo {author} {\bibfnamefont {J.}~\bibnamefont
  {Brehmer}}, \bibinfo {author} {\bibfnamefont {J.}~\bibnamefont {Hewett}},
  \bibinfo {author} {\bibfnamefont {J.}~\bibnamefont {Kopp}}, \bibinfo {author}
  {\bibfnamefont {T.}~\bibnamefont {Rizzo}}, \ and\ \bibinfo {author}
  {\bibfnamefont {J.}~\bibnamefont {Tattersall}},\ }\href {\doibase
  10.1007/JHEP10(2015)182} {\bibfield  {journal} {\bibinfo  {journal} {JHEP}\
  }\textbf {\bibinfo {volume} {10}},\ \bibinfo {pages} {182} (\bibinfo {year}
  {2015})},\ \Eprint {http://arxiv.org/abs/1507.00013} {arXiv:1507.00013
  [hep-ph]} \BibitemShut {NoStop}%
\bibitem [{\citenamefont {collaboration}(2015)}]{atlas}%
  \BibitemOpen
  \bibfield  {author} {\bibinfo {author} {\bibfnamefont {T.~A.}\ \bibnamefont
  {collaboration}} (\bibinfo {collaboration} {ATLAS}),\ }\href@noop {} {\emph
  {\bibinfo {title} {{Search for resonances decaying to photon pairs in 3.2
  fb$^{-1}$ of $pp$ collisions at $\sqrt{s}$ = 13 TeV with the ATLAS
  detector}}}},\ \bibinfo {type} {Tech. Rep.}\ \bibinfo {number}
  {ATLAS-CONF-2015-081}\ (\bibinfo {year} {2015})\BibitemShut {NoStop}%
\bibitem [{\citenamefont {Collaboration}(2015)}]{CMS:2015dxe}%
  \BibitemOpen
  \bibfield  {author} {\bibinfo {author} {\bibfnamefont {C.}~\bibnamefont
  {Collaboration}} (\bibinfo {collaboration} {CMS}),\ }\href@noop {} {\emph
  {\bibinfo {title} {{Search for new physics in high mass diphoton events in
  proton-proton collisions at 13TeV}}}},\ \bibinfo {type} {Tech. Rep.}\
  \bibinfo {number} {CMS-PAS-EXO-15-004}\ (\bibinfo {year} {2015})\BibitemShut
  {NoStop}%
\bibitem [{\citenamefont {Nakamura}\ \emph {et~al.}(2010)\citenamefont
  {Nakamura} \emph {et~al.}}]{Nakamura:2010zzi}%
  \BibitemOpen
  \bibfield  {author} {\bibinfo {author} {\bibfnamefont {K.}~\bibnamefont
  {Nakamura}} \emph {et~al.} (\bibinfo {collaboration} {Particle Data Group}),\
  }\href {\doibase 10.1088/0954-3899/37/7A/075021} {\bibfield  {journal}
  {\bibinfo  {journal} {J.Phys.G}\ }\textbf {\bibinfo {volume} {G37}},\
  \bibinfo {pages} {075021} (\bibinfo {year} {2010})}\BibitemShut {NoStop}%
\bibitem [{\citenamefont {Antusch}\ and\ \citenamefont
  {Fischer}(2014)}]{Antusch:2014woa}%
  \BibitemOpen
  \bibfield  {author} {\bibinfo {author} {\bibfnamefont {S.}~\bibnamefont
  {Antusch}}\ and\ \bibinfo {author} {\bibfnamefont {O.}~\bibnamefont
  {Fischer}},\ }\href {\doibase 10.1007/JHEP10(2014)094} {\bibfield  {journal}
  {\bibinfo  {journal} {JHEP}\ }\textbf {\bibinfo {volume} {1410}},\ \bibinfo
  {pages} {94} (\bibinfo {year} {2014})},\ \Eprint
  {http://arxiv.org/abs/1407.6607} {arXiv:1407.6607 [hep-ph]} \BibitemShut
  {NoStop}%
\bibitem [{\citenamefont {Antusch}\ \emph {et~al.}(2006)\citenamefont
  {Antusch}, \citenamefont {Biggio}, \citenamefont {Fernandez-Martinez},
  \citenamefont {Gavela},\ and\ \citenamefont {Lopez-Pavon}}]{Antusch:2006vwa}%
  \BibitemOpen
  \bibfield  {author} {\bibinfo {author} {\bibfnamefont {S.}~\bibnamefont
  {Antusch}}, \bibinfo {author} {\bibfnamefont {C.}~\bibnamefont {Biggio}},
  \bibinfo {author} {\bibfnamefont {E.}~\bibnamefont {Fernandez-Martinez}},
  \bibinfo {author} {\bibfnamefont {M.}~\bibnamefont {Gavela}}, \ and\ \bibinfo
  {author} {\bibfnamefont {J.}~\bibnamefont {Lopez-Pavon}},\ }\href {\doibase
  10.1088/1126-6708/2006/10/084} {\bibfield  {journal} {\bibinfo  {journal}
  {JHEP}\ }\textbf {\bibinfo {volume} {0610}},\ \bibinfo {pages} {084}
  (\bibinfo {year} {2006})},\ \Eprint {http://arxiv.org/abs/hep-ph/0607020}
  {arXiv:hep-ph/0607020 [hep-ph]} \BibitemShut {NoStop}%
\bibitem [{\citenamefont {Chatrchyan}\ \emph
  {et~al.}(2013{\natexlab{a}})\citenamefont {Chatrchyan} \emph
  {et~al.}}]{Chatrchyan:2012rva}%
  \BibitemOpen
  \bibfield  {author} {\bibinfo {author} {\bibfnamefont {S.}~\bibnamefont
  {Chatrchyan}} \emph {et~al.} (\bibinfo {collaboration} {CMS Collaboration}),\
  }\href {\doibase 10.1007/JHEP02(2013)036} {\bibfield  {journal} {\bibinfo
  {journal} {JHEP}\ }\textbf {\bibinfo {volume} {1302}},\ \bibinfo {pages}
  {036} (\bibinfo {year} {2013}{\natexlab{a}})},\ \Eprint
  {http://arxiv.org/abs/1211.5779} {arXiv:1211.5779 [hep-ex]} \BibitemShut
  {NoStop}%
\bibitem [{\citenamefont {Chatrchyan}\ \emph {et~al.}(2012)\citenamefont
  {Chatrchyan} \emph {et~al.}}]{CMS:2012xva}%
  \BibitemOpen
  \bibfield  {author} {\bibinfo {author} {\bibfnamefont {S.}~\bibnamefont
  {Chatrchyan}} \emph {et~al.} (\bibinfo {collaboration} {CMS}),\ }\href@noop
  {} {\emph {\bibinfo {title} {{Search for high-mass resonances decaying to top
  quark pairs in the lepton+jets channel}}}},\ \bibinfo {type} {Tech. Rep.}\
  \bibinfo {number} {CMS-PAS-EXO-11-093}\ (\bibinfo {address} {Geneva},\
  \bibinfo {year} {2012})\BibitemShut {NoStop}%
\bibitem [{\citenamefont {Aad}\ \emph {et~al.}(2013{\natexlab{a}})\citenamefont
  {Aad} \emph {et~al.}}]{Aad:2012raa}%
  \BibitemOpen
  \bibfield  {author} {\bibinfo {author} {\bibfnamefont {G.}~\bibnamefont
  {Aad}} \emph {et~al.} (\bibinfo {collaboration} {ATLAS Collaboration}),\
  }\href {\doibase 10.1007/JHEP01(2013)116} {\bibfield  {journal} {\bibinfo
  {journal} {JHEP}\ }\textbf {\bibinfo {volume} {1301}},\ \bibinfo {pages}
  {116} (\bibinfo {year} {2013}{\natexlab{a}})},\ \Eprint
  {http://arxiv.org/abs/1211.2202} {arXiv:1211.2202 [hep-ex]} \BibitemShut
  {NoStop}%
\bibitem [{\citenamefont {Aad}\ \emph {et~al.}(2013{\natexlab{b}})\citenamefont
  {Aad} \emph {et~al.}}]{Aad:2013wxa}%
  \BibitemOpen
  \bibfield  {author} {\bibinfo {author} {\bibfnamefont {G.}~\bibnamefont
  {Aad}} \emph {et~al.} (\bibinfo {collaboration} {ATLAS Collaboration}),\
  }\href@noop {} {\  (\bibinfo {year} {2013}{\natexlab{b}})},\ \Eprint
  {http://arxiv.org/abs/1305.0125} {arXiv:1305.0125 [hep-ex]} \BibitemShut
  {NoStop}%
\bibitem [{\citenamefont {Aad}\ \emph {et~al.}(2013{\natexlab{c}})\citenamefont
  {Aad} \emph {et~al.}}]{Aad:2013nca}%
  \BibitemOpen
  \bibfield  {author} {\bibinfo {author} {\bibfnamefont {G.}~\bibnamefont
  {Aad}} \emph {et~al.} (\bibinfo {collaboration} {ATLAS Collaboration}),\
  }\href@noop {} {\  (\bibinfo {year} {2013}{\natexlab{c}})},\ \Eprint
  {http://arxiv.org/abs/1305.2756} {arXiv:1305.2756 [hep-ex]} \BibitemShut
  {NoStop}%
\bibitem [{\citenamefont {Chatrchyan}\ \emph
  {et~al.}(2013{\natexlab{b}})\citenamefont {Chatrchyan} \emph
  {et~al.}}]{Chatrchyan:2013qha}%
  \BibitemOpen
  \bibfield  {author} {\bibinfo {author} {\bibfnamefont {S.}~\bibnamefont
  {Chatrchyan}} \emph {et~al.} (\bibinfo {collaboration} {CMS Collaboration}),\
  }\href@noop {} {\  (\bibinfo {year} {2013}{\natexlab{b}})},\ \Eprint
  {http://arxiv.org/abs/1302.4794} {arXiv:1302.4794 [hep-ex]} \BibitemShut
  {NoStop}%
\bibitem [{\citenamefont {Chatrchyan}\ \emph
  {et~al.}(2013{\natexlab{c}})\citenamefont {Chatrchyan} \emph
  {et~al.}}]{Chatrchyan:2012oaa}%
  \BibitemOpen
  \bibfield  {author} {\bibinfo {author} {\bibfnamefont {S.}~\bibnamefont
  {Chatrchyan}} \emph {et~al.} (\bibinfo {collaboration} {CMS Collaboration}),\
  }\href {\doibase 10.1016/j.physletb.2013.02.003} {\bibfield  {journal}
  {\bibinfo  {journal} {Phys.Lett.}\ }\textbf {\bibinfo {volume} {B720}},\
  \bibinfo {pages} {63} (\bibinfo {year} {2013}{\natexlab{c}})},\ \Eprint
  {http://arxiv.org/abs/1212.6175} {arXiv:1212.6175 [hep-ex]} \BibitemShut
  {NoStop}%
\bibitem [{\citenamefont {Aad}\ \emph {et~al.}(2012)\citenamefont {Aad} \emph
  {et~al.}}]{Aad:2012hf}%
  \BibitemOpen
  \bibfield  {author} {\bibinfo {author} {\bibfnamefont {G.}~\bibnamefont
  {Aad}} \emph {et~al.} (\bibinfo {collaboration} {ATLAS Collaboration}),\
  }\href {\doibase 10.1007/JHEP11(2012)138} {\bibfield  {journal} {\bibinfo
  {journal} {JHEP}\ }\textbf {\bibinfo {volume} {1211}},\ \bibinfo {pages}
  {138} (\bibinfo {year} {2012})},\ \Eprint {http://arxiv.org/abs/1209.2535}
  {arXiv:1209.2535 [hep-ex]} \BibitemShut {NoStop}%
\bibitem [{CMS(2013)}]{CMS-PAS-EXO-12-061}%
  \BibitemOpen
  \href@noop {} {\emph {\bibinfo {title} {{Search for Resonances in the
  Dilepton Mass Distribution in pp Collisions at sqrt(s) = 8 TeV}}}},\ \bibinfo
  {type} {Tech. Rep.}\ \bibinfo {number} {CMS-PAS-EXO-12-061}\ (\bibinfo
  {institution} {CERN},\ \bibinfo {address} {Geneva},\ \bibinfo {year}
  {2013})\BibitemShut {NoStop}%
\bibitem [{\citenamefont {{Chatrchyan}}\ \emph {et~al.}(2012)\citenamefont
  {{Chatrchyan}}, \citenamefont {{Khachatryan}}, \citenamefont {{Sirunyan}},
  \citenamefont {{Tumasyan}}, \citenamefont {{Adam}}, \citenamefont
  {{Bergauer}}, \citenamefont {{Dragicevic}}, \citenamefont {{Er{\"o}}},
  \citenamefont {{Fabjan}}, \citenamefont {{Friedl}},\ and\ \citenamefont
  {et~al.}}]{2012PhLB..714..158C}%
  \BibitemOpen
  \bibfield  {author} {\bibinfo {author} {\bibfnamefont {S.}~\bibnamefont
  {{Chatrchyan}}}, \bibinfo {author} {\bibfnamefont {V.}~\bibnamefont
  {{Khachatryan}}}, \bibinfo {author} {\bibfnamefont {A.~M.}\ \bibnamefont
  {{Sirunyan}}}, \bibinfo {author} {\bibfnamefont {A.}~\bibnamefont
  {{Tumasyan}}}, \bibinfo {author} {\bibfnamefont {W.}~\bibnamefont {{Adam}}},
  \bibinfo {author} {\bibfnamefont {T.}~\bibnamefont {{Bergauer}}}, \bibinfo
  {author} {\bibfnamefont {M.}~\bibnamefont {{Dragicevic}}}, \bibinfo {author}
  {\bibfnamefont {J.}~\bibnamefont {{Er{\"o}}}}, \bibinfo {author}
  {\bibfnamefont {C.}~\bibnamefont {{Fabjan}}}, \bibinfo {author}
  {\bibfnamefont {M.}~\bibnamefont {{Friedl}}}, \ and\ \bibinfo {author}
  {\bibnamefont {et~al.}},\ }\href {\doibase 10.1016/j.physletb.2012.06.051}
  {\bibfield  {journal} {\bibinfo  {journal} {Physics Letters B}\ }\textbf
  {\bibinfo {volume} {714}},\ \bibinfo {pages} {158} (\bibinfo {year}
  {2012})},\ \Eprint {http://arxiv.org/abs/1206.1849} {arXiv:1206.1849
  [hep-ex]} \BibitemShut {NoStop}%
\bibitem [{\citenamefont {Aad}\ \emph {et~al.}(2015{\natexlab{b}})\citenamefont
  {Aad} \emph {et~al.}}]{atlas13}%
  \BibitemOpen
  \bibfield  {author} {\bibinfo {author} {\bibfnamefont {G.}~\bibnamefont
  {Aad}} \emph {et~al.} (\bibinfo {collaboration} {ATLAS}),\ }\href@noop {}
  {\emph {\bibinfo {title} {{Search for new phenomena in the dilepton final
  state using proton-proton collisions at √ s = 13 TeV with the ATLAS
  detector}}}},\ \bibinfo {type} {Tech. Rep.}\ (\bibinfo {year}
  {2015})\BibitemShut {NoStop}%
\bibitem [{\citenamefont {Langacker}(2009)}]{Langacker:2008yv}%
  \BibitemOpen
  \bibfield  {author} {\bibinfo {author} {\bibfnamefont {P.}~\bibnamefont
  {Langacker}},\ }\href {\doibase 10.1103/RevModPhys.81.1199} {\bibfield
  {journal} {\bibinfo  {journal} {Rev.Mod.Phys.}\ }\textbf {\bibinfo {volume}
  {81}},\ \bibinfo {pages} {1199} (\bibinfo {year} {2009})},\ \Eprint
  {http://arxiv.org/abs/0801.1345} {arXiv:0801.1345 [hep-ph]} \BibitemShut
  {NoStop}%
\bibitem [{\citenamefont {Allanach}\ \emph {et~al.}(2015)\citenamefont
  {Allanach}, \citenamefont {Queiroz}, \citenamefont {Strumia},\ and\
  \citenamefont {Sun}}]{Allanach:2015gkd}%
  \BibitemOpen
  \bibfield  {author} {\bibinfo {author} {\bibfnamefont {B.}~\bibnamefont
  {Allanach}}, \bibinfo {author} {\bibfnamefont {F.~S.}\ \bibnamefont
  {Queiroz}}, \bibinfo {author} {\bibfnamefont {A.}~\bibnamefont {Strumia}}, \
  and\ \bibinfo {author} {\bibfnamefont {S.}~\bibnamefont {Sun}},\ }\href@noop
  {} {\  (\bibinfo {year} {2015})},\ \Eprint {http://arxiv.org/abs/1511.07447}
  {arXiv:1511.07447 [hep-ph]} \BibitemShut {NoStop}%
\bibitem [{\citenamefont {Aad}\ \emph {et~al.}(2014{\natexlab{a}})\citenamefont
  {Aad} \emph {et~al.}}]{Aad:2014qaa}%
  \BibitemOpen
  \bibfield  {author} {\bibinfo {author} {\bibfnamefont {G.}~\bibnamefont
  {Aad}} \emph {et~al.} (\bibinfo {collaboration} {ATLAS Collaboration}),\
  }\href {\doibase 10.1007/JHEP06(2014)124} {\bibfield  {journal} {\bibinfo
  {journal} {JHEP}\ }\textbf {\bibinfo {volume} {1406}},\ \bibinfo {pages}
  {124} (\bibinfo {year} {2014}{\natexlab{a}})},\ \Eprint
  {http://arxiv.org/abs/1403.4853} {arXiv:1403.4853 [hep-ex]} \BibitemShut
  {NoStop}%
\bibitem [{\citenamefont {Aad}\ \emph {et~al.}(2014{\natexlab{b}})\citenamefont
  {Aad} \emph {et~al.}}]{Aad:2014vma}%
  \BibitemOpen
  \bibfield  {author} {\bibinfo {author} {\bibfnamefont {G.}~\bibnamefont
  {Aad}} \emph {et~al.} (\bibinfo {collaboration} {ATLAS Collaboration}),\
  }\href {\doibase 10.1007/JHEP05(2014)071} {\bibfield  {journal} {\bibinfo
  {journal} {JHEP}\ }\textbf {\bibinfo {volume} {1405}},\ \bibinfo {pages}
  {071} (\bibinfo {year} {2014}{\natexlab{b}})},\ \Eprint
  {http://arxiv.org/abs/1403.5294} {arXiv:1403.5294 [hep-ex]} \BibitemShut
  {NoStop}%
\bibitem [{\citenamefont {Aad}\ \emph {et~al.}(2014{\natexlab{c}})\citenamefont
  {Aad} \emph {et~al.}}]{Aad:2014iza}%
  \BibitemOpen
  \bibfield  {author} {\bibinfo {author} {\bibfnamefont {G.}~\bibnamefont
  {Aad}} \emph {et~al.} (\bibinfo {collaboration} {ATLAS Collaboration}),\
  }\href {\doibase 10.1103/PhysRevD.90.052001} {\bibfield  {journal} {\bibinfo
  {journal} {Phys.Rev.}\ }\textbf {\bibinfo {volume} {D90}},\ \bibinfo {pages}
  {052001} (\bibinfo {year} {2014}{\natexlab{c}})},\ \Eprint
  {http://arxiv.org/abs/1405.5086} {arXiv:1405.5086 [hep-ex]} \BibitemShut
  {NoStop}%
\bibitem [{\citenamefont {Chen}\ \emph {et~al.}(2015)\citenamefont {Chen},
  \citenamefont {Cheng},\ and\ \citenamefont {Low}}]{Chen:2015jmn}%
  \BibitemOpen
  \bibfield  {author} {\bibinfo {author} {\bibfnamefont {C.-R.}\ \bibnamefont
  {Chen}}, \bibinfo {author} {\bibfnamefont {H.-C.}\ \bibnamefont {Cheng}}, \
  and\ \bibinfo {author} {\bibfnamefont {I.}~\bibnamefont {Low}},\ }\href@noop
  {} {\  (\bibinfo {year} {2015})},\ \Eprint {http://arxiv.org/abs/1511.01452}
  {arXiv:1511.01452 [hep-ph]} \BibitemShut {NoStop}%
\bibitem [{\citenamefont {Berger}\ \emph {et~al.}(2014)\citenamefont {Berger},
  \citenamefont {Giddings}, \citenamefont {Wang},\ and\ \citenamefont
  {Zhang}}]{Berger:2014gga}%
  \BibitemOpen
  \bibfield  {author} {\bibinfo {author} {\bibfnamefont {E.~L.}\ \bibnamefont
  {Berger}}, \bibinfo {author} {\bibfnamefont {S.~B.}\ \bibnamefont
  {Giddings}}, \bibinfo {author} {\bibfnamefont {H.}~\bibnamefont {Wang}}, \
  and\ \bibinfo {author} {\bibfnamefont {H.}~\bibnamefont {Zhang}},\ }\href
  {\doibase 10.1103/PhysRevD.90.076004} {\bibfield  {journal} {\bibinfo
  {journal} {Phys. Rev.}\ }\textbf {\bibinfo {volume} {D90}},\ \bibinfo {pages}
  {076004} (\bibinfo {year} {2014})},\ \Eprint {http://arxiv.org/abs/1406.6054}
  {arXiv:1406.6054 [hep-ph]} \BibitemShut {NoStop}%
\bibitem [{\citenamefont {Curtin}\ \emph {et~al.}(2013)\citenamefont {Curtin},
  \citenamefont {Essig}, \citenamefont {Gori}, \citenamefont {Jaiswal},
  \citenamefont {Katz} \emph {et~al.}}]{Curtin:2013fra}%
  \BibitemOpen
  \bibfield  {author} {\bibinfo {author} {\bibfnamefont {D.}~\bibnamefont
  {Curtin}}, \bibinfo {author} {\bibfnamefont {R.}~\bibnamefont {Essig}},
  \bibinfo {author} {\bibfnamefont {S.}~\bibnamefont {Gori}}, \bibinfo {author}
  {\bibfnamefont {P.}~\bibnamefont {Jaiswal}}, \bibinfo {author} {\bibfnamefont
  {A.}~\bibnamefont {Katz}},  \emph {et~al.},\ }\href@noop {} {\  (\bibinfo
  {year} {2013})},\ \Eprint {http://arxiv.org/abs/1312.4992} {arXiv:1312.4992
  [hep-ph]} \BibitemShut {NoStop}%
\bibitem [{\citenamefont {Huang}\ \emph {et~al.}(2014)\citenamefont {Huang},
  \citenamefont {Liu}, \citenamefont {Wang},\ and\ \citenamefont
  {Yu}}]{Huang:2013ima}%
  \BibitemOpen
  \bibfield  {author} {\bibinfo {author} {\bibfnamefont {J.}~\bibnamefont
  {Huang}}, \bibinfo {author} {\bibfnamefont {T.}~\bibnamefont {Liu}}, \bibinfo
  {author} {\bibfnamefont {L.-T.}\ \bibnamefont {Wang}}, \ and\ \bibinfo
  {author} {\bibfnamefont {F.}~\bibnamefont {Yu}},\ }\href {\doibase
  10.1103/PhysRevLett.112.221803} {\bibfield  {journal} {\bibinfo  {journal}
  {Phys.Rev.Lett.}\ }\textbf {\bibinfo {volume} {112}},\ \bibinfo {pages}
  {221803} (\bibinfo {year} {2014})},\ \Eprint {http://arxiv.org/abs/1309.6633}
  {arXiv:1309.6633 [hep-ph]} \BibitemShut {NoStop}%
\bibitem [{\citenamefont {Moreau}(2013)}]{Moreau:2012da}%
  \BibitemOpen
  \bibfield  {author} {\bibinfo {author} {\bibfnamefont {G.}~\bibnamefont
  {Moreau}},\ }\href {\doibase 10.1103/PhysRevD.87.015027} {\bibfield
  {journal} {\bibinfo  {journal} {Phys. Rev.}\ }\textbf {\bibinfo {volume}
  {D87}},\ \bibinfo {pages} {015027} (\bibinfo {year} {2013})},\ \Eprint
  {http://arxiv.org/abs/1210.3977} {arXiv:1210.3977 [hep-ph]} \BibitemShut
  {NoStop}%
\bibitem [{\citenamefont {Angelescu}\ \emph {et~al.}(2015)\citenamefont
  {Angelescu}, \citenamefont {Djouadi},\ and\ \citenamefont
  {Moreau}}]{Angelescu:2015kga}%
  \BibitemOpen
  \bibfield  {author} {\bibinfo {author} {\bibfnamefont {A.}~\bibnamefont
  {Angelescu}}, \bibinfo {author} {\bibfnamefont {A.}~\bibnamefont {Djouadi}},
  \ and\ \bibinfo {author} {\bibfnamefont {G.}~\bibnamefont {Moreau}},\
  }\href@noop {} {\  (\bibinfo {year} {2015})},\ \Eprint
  {http://arxiv.org/abs/1510.07527} {arXiv:1510.07527 [hep-ph]} \BibitemShut
  {NoStop}%
\bibitem [{\citenamefont {Anandakrishnan}\ \emph {et~al.}(2015)\citenamefont
  {Anandakrishnan}, \citenamefont {Collins}, \citenamefont {Farina},
  \citenamefont {Kuflik},\ and\ \citenamefont
  {Perelstein}}]{Anandakrishnan:2015yfa}%
  \BibitemOpen
  \bibfield  {author} {\bibinfo {author} {\bibfnamefont {A.}~\bibnamefont
  {Anandakrishnan}}, \bibinfo {author} {\bibfnamefont {J.~H.}\ \bibnamefont
  {Collins}}, \bibinfo {author} {\bibfnamefont {M.}~\bibnamefont {Farina}},
  \bibinfo {author} {\bibfnamefont {E.}~\bibnamefont {Kuflik}}, \ and\ \bibinfo
  {author} {\bibfnamefont {M.}~\bibnamefont {Perelstein}},\ }\href@noop {} {\
  (\bibinfo {year} {2015})},\ \Eprint {http://arxiv.org/abs/1506.05130}
  {arXiv:1506.05130 [hep-ph]} \BibitemShut {NoStop}%
\bibitem [{\citenamefont {Berger}\ \emph {et~al.}(2012)\citenamefont {Berger},
  \citenamefont {Hubisz},\ and\ \citenamefont {Perelstein}}]{Berger:2012ec}%
  \BibitemOpen
  \bibfield  {author} {\bibinfo {author} {\bibfnamefont {J.}~\bibnamefont
  {Berger}}, \bibinfo {author} {\bibfnamefont {J.}~\bibnamefont {Hubisz}}, \
  and\ \bibinfo {author} {\bibfnamefont {M.}~\bibnamefont {Perelstein}},\
  }\href {\doibase 10.1007/JHEP07(2012)016} {\bibfield  {journal} {\bibinfo
  {journal} {JHEP}\ }\textbf {\bibinfo {volume} {07}},\ \bibinfo {pages} {016}
  (\bibinfo {year} {2012})},\ \Eprint {http://arxiv.org/abs/1205.0013}
  {arXiv:1205.0013 [hep-ph]} \BibitemShut {NoStop}%
\end{thebibliography}%

\end{document}